\documentclass[manuscript]{aastex}

\usepackage{graphicx,amsmath,booktabs,threeparttable,url}
\usepackage{hyperref}

\usepackage{threeparttable}

\usepackage[usenames,dvips]{color}

\usepackage[normalem]{ulem}

\usepackage{natbib}

\begin{document}

\title{EVOLUTION OF POST-IMPACT COMPANION STARS IN SN~Ia REMNANTS WITHIN THE SINGLE-DEGENERATE SCENARIO }

\author{Kuo-Chuan Pan$^1$, Paul M. Ricker$^1$, and Ronald E. Taam$^{2,3}$}
\affil{$^1$Department of Astronomy, University of Illinois at Urbana$-$Champaign, 1002 West Green Street, Urbana, IL 61801, USA; kpan2@illinois.edu, pmricker@illinois.edu}
\affil{$^2$Department of Physics and Astronomy, Northwestern University, 2145 Sheridan Road, Evanston, IL 60208, USA; r-taam@northwestern.edu}
\affil{$^3$Academia Sinica Institute of Astronomy and Astrophysics, P.O. Box 23-141, Taipei 10617, Taiwan}

\begin{abstract}

The nature of the progenitor systems of Type Ia supernovae is still uncertain.  
One way to distinguish between the single-degenerate scenario (SDS) and 
double-degenerate scenario (DDS) is to search for the post-impact remnant star.  To examine the 
characteristics of the post-impact remnant star, we have carried out three-dimensional hydrodynamic 
simulations of supernova impacts on main sequence-like stars. 
We explore the evolution of the post-impact remnants using the stellar evolution code MESA. 
We find that the luminosity and radius of the remnant star dramatically increase just after the impact. 
After the explosion, post-impact companions continue to expand on a progenitor-dependent timescale of 
$\sim 10^{2.5-3} $~yr before contracting. 
It is found that the time evolution of the remnant star is 
dependent not only on the amount of energy absorbed, but also on the depth of the energy deposition. 
We examine the viability of the candidate star Tycho G as the possible remnant companion in Tycho's supernova by comparing it to the evolved post-impact remnant stars in our simulations.
The closest model in our simulations has a similar effective temperature, 
but the luminosity and radius are twice as large.
By examining the angular momentum distribution in our simulations,
we find that the surface rotational speed could drop to $\sim 10$~km~s$^{-1}$ 
if the specific angular momentum is conserved during the post-impact evolution, 
implying that Tycho G cannot be completely ruled out because of its low surface rotation speed. 

\end{abstract}



\section{INTRODUCTION}

Type Ia supernovae (SNe~Ia) are important as cosmological distance probes and sources of metal enrichment for the interstellar medium.
Although they can be treated as ``standardizable candles,'' intrinsic variations in SN~Ia properties do exist \citep{2000tias.conf...33L, 2000ARA&A..38..191H}.
In principle these could be associated with different SN~Ia progenitor systems and explosion mechanisms. 
It is widely accepted that SNe~Ia are thermonuclear explosions of carbon-oxygen (CO) white dwarfs (WDs), 
but so far observations have not conclusively favored only a single model \citep{2000tias.conf...33L, 2000ARA&A..38..191H,2012Natur.481..149R, 2012NewAR..56..122W}.

The mainstream progenitor scenarios include the single-degenerate scenario 
(SDS; \cite{1973ApJ...186.1007W, 1982ApJ...257..780N})
and the double-degenerate scenario 
(DDS; \cite{1984ApJS...54..335I, 1984ApJ...277..355W}). 
The SDS involves a CO WD accreting mass from a non-degenerate companion,
such as a main sequence (MS) star, red giant (RG), or helium (He) star, through Roche-lobe overflow (RLOF). 
Once the CO WD reaches the Chandrasekhar mass, 
it becomes unstable and explodes as an SN~Ia. 
The DDS instead involves gravitational wave-driven merging of two CO WDs
with total mass greater than the Chandrasekhar mass. 
Recent observations of the nearby SN~2011fe provide useful constraints on its progenitor system \citep{2011Natur.480..348L,2011Natur.480..344N,2012ApJ...744L..17B,2012ApJ...749...18B,2012ApJ...750..164C}, though both scenarios may contribute to the SN~Ia population.

To distinguish between the SDS and DDS, different approaches have been proposed during the past decade.
For example, detection of the pre-supernova circumstellar medium around supernova remnants (SNRs) provides evidence of the mass transfer phase in the SDS \citep{1995PASP..107.1019B,2011Sci...333..856S,2012ApJ...752..101F,2012Sci...337..942D}. 
Collision of the SN ejecta with a non-degenerate companion may affect the early SN light curve, depending on the companion type and viewing angle \citep{2010ApJ...708.1025K}.
Numerical simulations of the SDS also suggest that
the non-degenerate companion will survive the SN impact and should be detectable in the SNR \citep{2000ApJS..128..615M,2008A&A...489..943P, 2010ApJ...715...78P,2012ApJ...750..151P}.

\cite{2004Natur.431.1069R} found a star, which they called Tycho G, that is located in the central region of Tycho's SN and has a high proper motion, which in the SDS would be associated with the original orbital speed. This suggests that Tycho G could be a post-impact SDS remnant star.  
However, \cite{2009ApJ...701.1665K} found that the upper limit of the rotational speed of Tycho G is $\lesssim 7.5$~km~s$^{-1}$, which would seem to be a problem for the SDS interpretation,
since the companion stars in close binary systems are usually rapidly rotating.
\cite{2012ApJ...750..151P} address this problem by analyzing the angular momentum lost during the supernova impact, but the final rotational speed depends on the post-impact evolution of the remnant star.
Although many detailed studies of the effects of binary evolution on non-degenerate companions have been published \citep{1999ApJ...522..487H,2004MNRAS.350.1301H,2008ApJ...677L.109H,2008ApJ...679.1390H,2009MNRAS.395..847W,2009MNRAS.395.2103M,2010ApJ...710.1310M,2010A&A...515A..88W}, usually they have not included the SN impact and post-impact evolution.  

\cite{2003astro.ph..3660P} studied the post-impact evolution of a $1M_\odot$ subgiant by assuming a fixed amount of energy input and mass stripping. 
Unsurprisingly, they found that the post-impact star could be overluminous or underluminous compared to a normal star of the same temperature, 
depending on the choices of these parameters.
A more recent study by \cite{2012arXiv1205.5028S} examined the post-impact evolution of a $1M_\odot$ MS star from the hydrogen cataclysmic variables (HCV) scenario in \cite{2000ApJS..128..615M}, using an approach similar to that adopted by \cite{2003astro.ph..3660P}. 
They find an overluminous post-impact companion ($10-10^3 {\rm L}_\odot$) and
suggest that because of this Tycho G may not be associated with Tycho's SN.   
However, without detailed study of the impact of the SN ejecta on the companion star, especially 
shock compression in the stellar interior and the depth of energy deposition, these results 
cannot be considered definitive. 
Furthermore, the stellar mass and delay time for Tycho G might differ from the progenitor in the HCV scenario, which is the only model considered in \cite{2012arXiv1205.5028S}.

In this paper, we consider a wide range of progenitor models from the supersoft channel (WD+MS or WD+subgiant) in \cite{2008ApJ...679.1390H}(hereafter HKN) and perform three-dimensional hydrodynamics simulations of the impact of SN~Ia ejecta using the method described in \cite{2012ApJ...750..151P} (hereafter PRT).
We then use a one-dimensional stellar evolution code to study the post-impact evolution for each of these progenitor models. In the next section, the numerical methods and progenitor models are described. 
In Section~\ref{sec_evolution}, we present the results of post-impact evolution for different companion models and compare their observational quantities with Tycho G. 
In the final section, we summarize our results and conclude.


\section{NUMERICAL METHODS}
Our numerical simulations are divided into three stages.
The first stage uses a one-dimensional stellar evolution code to construct the progenitor system at the onset of the SN~Ia based on detailed binary evolution. 
The second stage uses a three-dimensional hydrodynamics code to simulate the impact of SN~Ia ejecta on the binary companion.
The final stage takes the post-impact companion star from the hydrodynamics simulation as input to the stellar evolution code, which is then used to
simulate the post-impact evolution.    


\subsection{Simulation Codes}
The hydrodynamics code used is FLASH\footnote{\url{http://flash.uchicago.edu}} version~3 \citep{2000ApJS..131..273F, 2008PhST..132a4046D}.
FLASH is a parallel, multi-dimensional hydrodynamics code based on block-structured adaptive mesh refinement (AMR). 
The equation of state (EOS) applied is the Helmholtz EOS \citep{2000ApJS..126..501T}.
The Poisson equation is solved on the AMR mesh using a Fourier transform-based multigrid algorithm \citep{2008ApJS..176..293R}.
The detailed numerical setup and initial conditions are the same as in PRT. 
In this paper, we consider only three-dimensional simulations that include orbital motion. 
The initial binary systems are assumed to be in RLOF, and the SN model used is the W7 model \citep{1984ApJ...286..644N}.

To construct the progenitor systems and simulate the post-impact evolution of the remnant star,
the stellar evolution code MESA\footnote{\url{http://mesa.sourceforge.net}} (Modules for Experiments in Stellar Astrophysics; \cite{2011ApJS..192....3P})
is employed.
The mixing-length theory (MLT) of convection used is the modified MLT by \cite{1965ApJ...142..841H}, which allows the convective efficiency
to vary with the opacity 
and has important effects on convective zones near the surfaces of stars. 
The ratio of mixing length to the local pressure scale height is set to $\alpha=1.918$, which is the value calibrated using the Sun.
The initial metallicity is assumed to be $Z=0.02$ for all the progenitor models.

\subsection{Progenitor Systems}
The progenitor systems are taken from the models in HKN.
HKN studied the binary evolution of WD+MS systems 
and found the region of the donor mass-orbital period plane for which SNe~Ia may occur.
In addition to RLOF mass transfer, HKN consider the optically thick wind from the WD and the mass stripping effects of a massive circumstellar torus. 
Although recent EVLA observations of SN~2011fe disagree with this scenario \citep{2012ApJ...750..164C}, HKN can explain the circumstellar matter
around SN~2002ic, SN~2005gj, and SN~2006X. 
Furthermore, \cite{2008ApJ...683L.127H} found that the theoretical delay-time distribution (DTD) based on this scenario is consistent with the observed DTD.

\begin{figure}
\epsscale{1.0}
\plotone{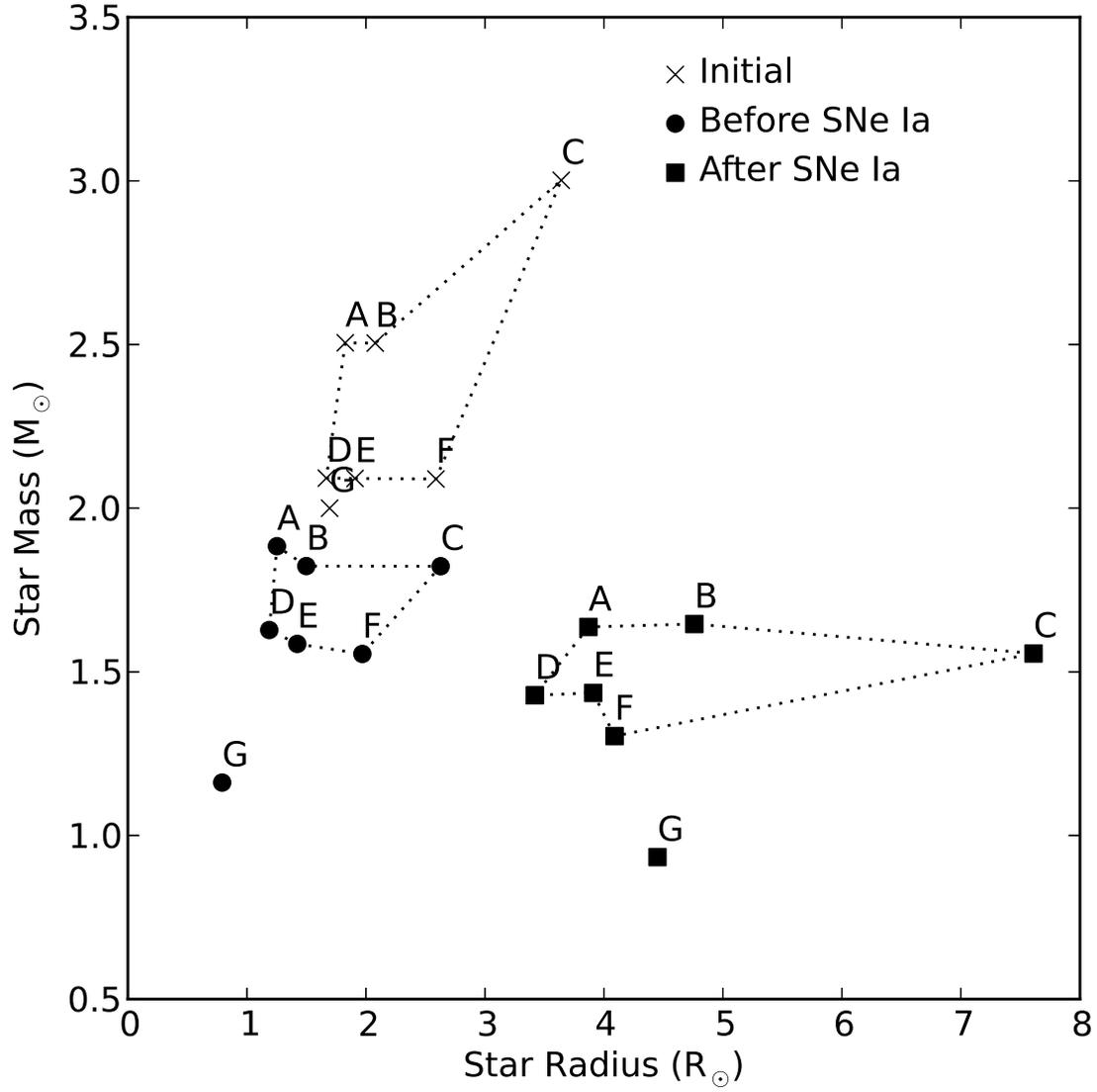}
\caption{\label{fig_rm} Stellar mass and radius for all considered progenitor models at different stages: initial state ($\times$),
just prior to SN (filled circles), after reaching hydrostatic equilibrium following SN (filled squares). }
\end{figure}

Figure~7 and Figure~8 in HKN show the initial and final regions of the donor mass-orbital period plane for WD+MS systems with initial WD mass
 $M_{\rm WD}= 1.0 M_\odot$, 
hydrogen composition $X=0.70$, metallicity $Z=0.02$, and stripping rate parameter $c_1=3$.  
We take six evolved MS models from these figures as our progenitor models (see Table~\ref{tab_models}).
Since these models are assumed to be in RLOF, we use MESA to evolve a zero-age-main-sequence (ZAMS) star with the initial mass and metallicity
taken from HKN until the radius of the MS star reaches its Roche-lobe radius.  
The Roche-lobe radius, $R_L$, is approximately given by \cite{1983ApJ...268..368E}:
\begin{equation}
\frac{R_{\rm L}}{a}=\frac{0.49 q^{2/3}}{0.6q^{2/3}+\ln(1+q^{1/3})}\ ,
\end{equation}
where $q$ is the mass ratio, and $a$ is the binary separation.
The binary separation can be found using Kepler's third law with the given mass and period from HKN. 
Although MESA has the ability to model binary evolution, we do not include the detailed physics of binary evolution as HKN considered but rather assume the MS star to have a constant mass-loss rate. 
We vary the mass-loss rate until the MS star matches the final mass and radius given in HKN within a timescale comparable to theirs ($\sim 10^6$ yr). 
The mass and radius of these MS stars are shown in Figure~\ref{fig_rm}.
Since we assume RLOF, the ones with longer periods have larger radii and longer delay times.

\begin{figure}
\plotone{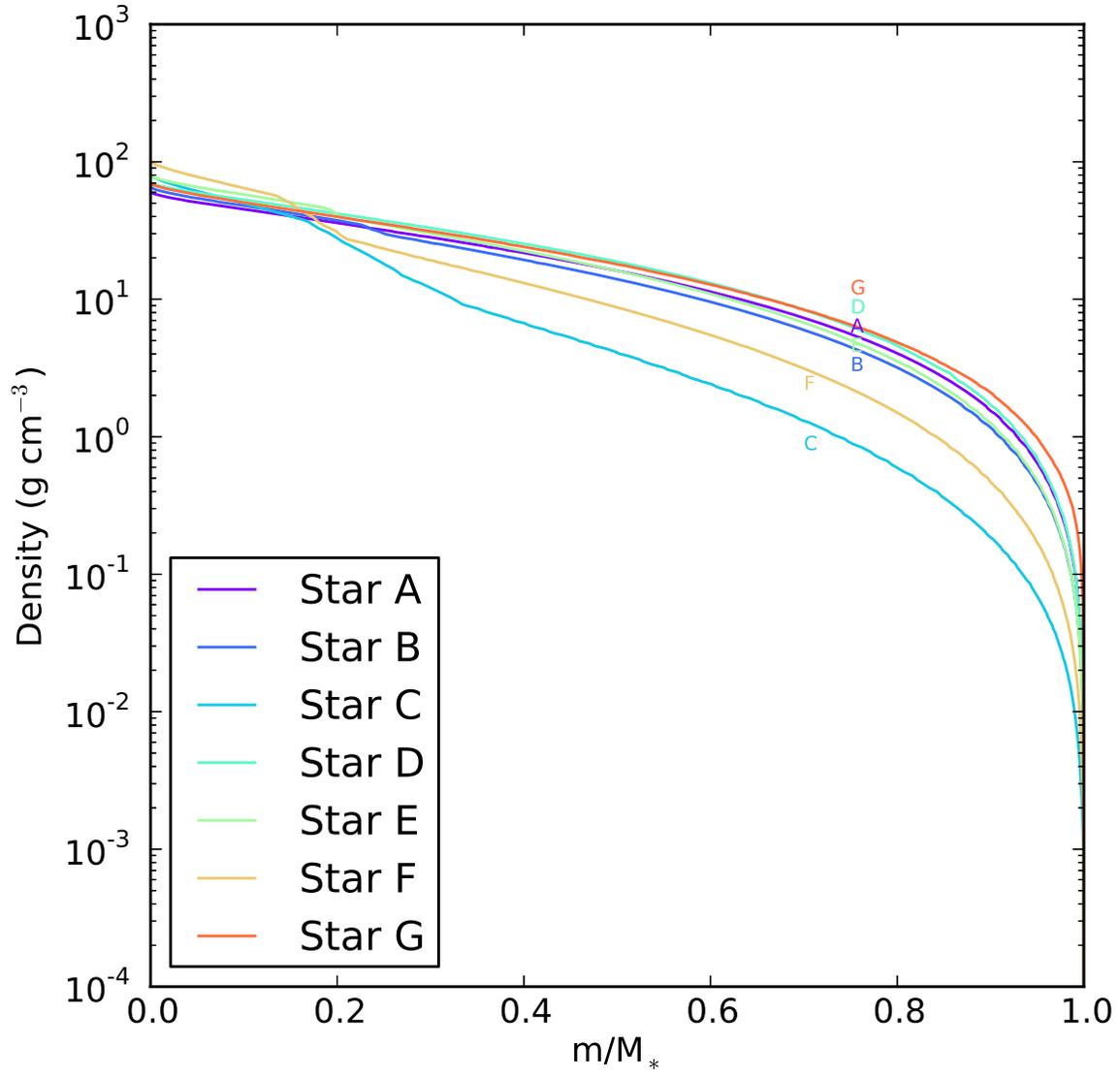}
\caption{\label{fig_models} Density vs. radius profiles for all the MS models at the time of the SN explosion.
The label for each model refers to Table~\ref{tab_models}.}
\end{figure}

Table~\ref{tab_models} summarizes the initial and final conditions of these six models we have created using MESA. 
The ``MS-r'' case in PRT is also included as a comparison (the star G in Table~\ref{tab_models}).  
Note that the HCV model in \cite{2000ApJS..128..615M, 2012arXiv1205.5028S} 
and the HCV' model in PRT are not considered here, 
since these scenarios did not consider detailed binary evolution.
Figure~\ref{fig_models} shows the density profiles of all these models at the time of the explosion. 
Most models are slightly evolved MS stars, but some of them are close to ZAMS, depending on the initial periods in HKN.  
MESA provides all the thermodynamical quantities we need for FLASH simulations. 
To save computation time, the compositions of the companion models are adjusted to $^1$H, $^4$He, plus $^{12}$C, 
where $^{12}$C represents all elements heavier than helium.
After converting MESA models to FLASH models, we perform FLASH simulations using the method described in PRT. 

\begin{table}
\begin{center}
\caption{The progenitor models \label{tab_models}}
\begin{tabular}{ccccccccccc}
\\
\tableline
Model & $M_{\rm i}$ \tablenotemark{a} & $P_{\rm i}$ & $R_{\rm i}$  & $L_{\rm i}$& $T_{\rm eff,i}$ & $M_{\rm f}$ \tablenotemark{b} & $P_{\rm f}$ & $R_{\rm f}$  & $L_{\rm f}$  & $T_{\rm eff,f}$ \\
 & ($M_\odot$) &  (day) & ($R_\odot$) &  ($L_\odot$) &  (K) & ($M_\odot$) &  (day) & ($R_\odot$) &  ($L_\odot$) &  (K)\\
\tableline
A & 2.51 & 0.477 & 1.83 & 39.2 & 10,696 &1.88 & 0.350 &1.25 & 2.35 & 6,392 \\
B & 2.51 & 0.600 & 2.08 & 42.4 & 10,224 & 1.92 & 0.466 & 1.50 & 3.64 & 6,516 \\
C & 3.01 & 1.23 & 3.64 & 110.0 & 9,800 & 1.82 & 1.09 & 2.63 & 8.06 & 6,003 \\
D & 2.09 & 0.472 & 1.67 & 19.2 & 9,358 & 1.63 & 0.353 &1.19 & 2.09 & 6,372 \\
E & 2.09 & 0.589 & 1.91 & 20.8 & 8,933 & 1.59 & 0.470 &1.42 & 3.15 & 6,450 \\	
F & 2.09 & 0.936 & 2.59 & 23.9 & 7,934 & 1.55 & 0.770 &1.97 & 5.09 & 6,182 \\
G\tablenotemark{*} & 2.00 & 1.00 & 1.70 & 17.6 & 9,083 & 1.17 & 0.233 & 0.792 & 0.463 & 5,355\\
\end{tabular}
\tablenotetext{a}{The mass ($M_{\rm i}$), period (day), radius ($R_{\rm i}$), luminosity ($L_{\rm i}$), and effective temperature ($T_{\rm eff,i}$)
for different progenitor models at the beginning of RLOF for WD+MS systems, using the initial masses and orbital periods in Figure~7 of HKN.}
\tablenotetext{b}{The mass ($M_{\rm f}$), period (day), radius ($R_{\rm f}$), luminosity ($L_{\rm f}$), and effective temperature ($T_{\rm eff,f}$)
for different progenitor models at the time of the SN explosion, using the final masses and orbital periods in Figure~8 of HKN.}
\tablenotetext{*}{The MS-r model in PRT.}
\end{center}
\end{table}

\subsection{Post-Impact Companion Models}

\begin{table}
\begin{center}
\caption{Post-impact remnant stars \label{tab_post_models}}
\begin{tabular}{cccrccc}
\\
\tableline
Model  & M$_{\rm SN}$ ($M_\odot$) \tablenotemark{a}  & $R_{\rm SN}$ ($R_\odot$) & $\Delta M/ M_*$& $v_{\rm Linear}$ (km~s$^{-1}$) \tablenotemark{b} & $L_{\rm SN}$ ($L_\odot$) & $T_{\rm eff,SN}$ (K) \\
\tableline  
A & 1.64 & 3.87 & 12.8  \%& 179.0 & 16.9 & 5,954\\
B & 1.65 & 4.76 & 14.1  \%& 178.7 & 22.4 & 5,760\\
C & 1.56 & 7.61 & 14.3  \%& 136.3 & 44.2 & 5,398\\
D & 1.43 & 3.42 & 12.3  \%& 187.7 & 13.1 & 5,936\\
E & 1.44 & 3.91 & 9.4  \% & 190.9 & 14.6 & 5,715\\
F & 1.30 & 4.09 & 16.1 \%& 142.6 & 15.0 & 5,618\\
G & 0.93 & 4.45 & 20.1 \%& 271.0 & 13.9 & 5,289\\  
\end{tabular}
\tablenotetext{a}{The mass ($M_{\rm SN}$), radius ($R_{\rm SN}$), percentage of unbound mass ($\Delta M/ M_*$), linear velocity  ($v_{\rm Linear}$),
luminosity ($L_{\rm SN}$), and effective temperature ($T_{\rm eff,SN}$) of initial post-impact hydrostatic models in MESA.}
\tablenotetext{b}{\rm Linear velocity includes the pre-supernova orbital speed and kick velocity.}
\end{center}
\end{table}

\begin{figure}
\plotone{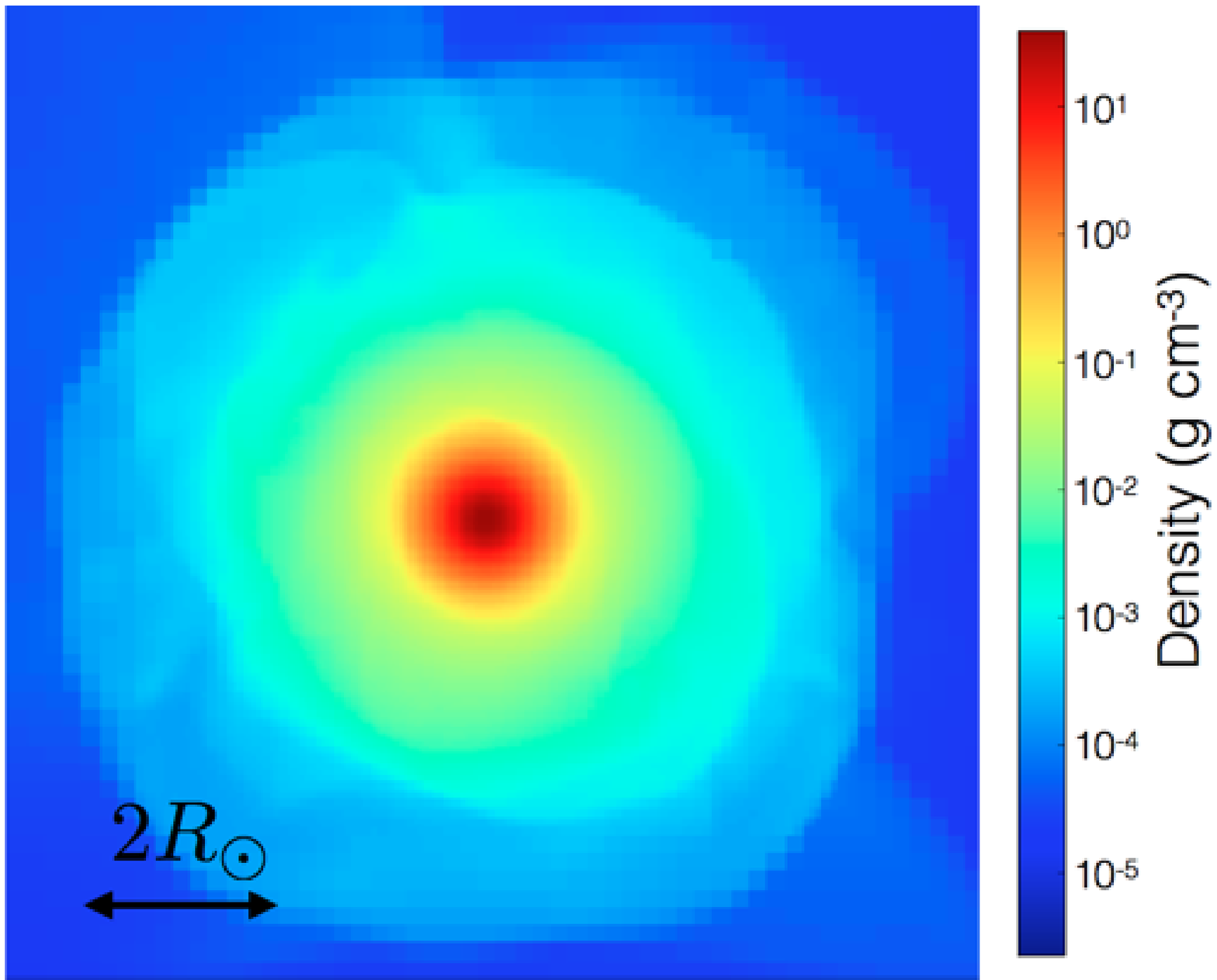}
\caption{\label{fig_dens} 
Gas density distribution in the orbital plane for a three-dimensional SN~Ia simulation with model F (see Table~\ref{tab_models}) 
at time $t=3.26 \times 10^4$ sec after the explosion. 
The frame shows a domain spanning 10$R_\odot$. The color scale indicates the logarithm of the gas density.}
\end{figure}

After the SN impact, the companion star is heated 
and loses about $\sim 10-20\%$ of its mass due to stripping and ablation by the SN ejecta 
(see PRT for detailed description of the SN impact).   
The remaining mass and final conditions of the remnant stars are summarized in Table~\ref{tab_post_models}.
Figure~\ref{fig_dens} shows a typical gas density distribution for model F in the orbital plane at the end of the simulation ($t=3.26 \times 10^4$~sec). 
It is difficult to run hydrodynamics simulations up to the age of historical SNRs ($\sim 500-1000$~yrs)
because the dynamical time scale, which determines the timestep, is only on the order of hundreds of seconds.
Moreover, the star is close to being spherically symmetric.
Therefore, the most straightforward method is to turn the post-impact results back into a one-dimensional problem and solve for the subsequent evolution
using MESA. 

To run simulations of the post-impact evolution of remnant stars in MESA, the first step is to convert the three-dimensional data into 
angle-averaged one-dimensional radial profiles.  Our setup in MESA cannot deal with the spin of the remnant star, and the angular momentum 
is finite at the end of the FLASH simulations.  However, PRT demonstrated that the loss of angular momentum during the SN impact will 
significantly decrease the spin.  Since the rotational velocity is significantly less than breakup, $\Omega_{\rm K}= (GM/R^3_*)^{1/2} \sim 5 \times 10^{-5}$~s$^{-1}$, we
ignore the spin during MESA simulations, but consider it separately using the post-impact specific 
angular momentum from FLASH.  The role of rotation will be discussed in more detail in Section\ref{sec_rotation}.

Since the companion models at the end of the FLASH simulations are not yet in hydrostatic equilibrium, 
the averaged radial profiles of density, temperature and pressure cannot be used directly in MESA.
However, the averaged specific entropy and composition mass profiles are conserved if the system is adiabatic. 
Since the Mach numbers at the end of the FLASH simulations are subsonic for most gravitationally bound gas, 
we can use these profiles, as a start, to construct hydrostatic models for MESA. 
Hydrostatic equilibrium should
be achieved relatively quickly after the SN impact in comparison with the thermal timescale.

To calculate the averaged specific entropy and composition profiles, 
only gravitationally bound gas is considered, and the center of the remnant star is taken to be the location of the gravitational potential minimum.
The simulation domain is then divided into $128-256$ radial bins, depending on the zone spacing
and the distance between the star's center and the boundary of the simulation box at the end of the FLASH simulation.  
Although FLASH and MESA include a more realistic EOS, for simplicity only an ``ideal gas plus radiation'' EOS is used to calculate the entropy profile.
Therefore the specific entropy is
\begin{equation}
S = \frac{N_A k_B}{\mu} \ln{\left(\frac{T^{3/2}}{\rho} \right)}+\frac{4aT^3}{3\rho}\ , \label{eq_entropy}
\end{equation}
where $N_A$ is the Avogadro constant, $k_B$ is the Boltzmann constant, $a$ is the radiation constant, $\rho$ is the bound gas density,
$T$ is the temperature, and $\mu$ is the mean molecular weight.  
Using the total bound mass ($M_*$), the EOS, and the entropy and composition profiles,
 the density, pressure ($P$), temperature ($T$), and radius can be calculated by solving the hydrostatic equations,
\begin{equation}
\frac{dr}{dM_r}=\frac{1}{4\pi r^2\rho},
\end{equation}
\begin{equation}
\frac{dP}{dM_r}=-\frac{GM_r}{4\pi r^4},
\end{equation}
where $G$ is the gravitational constant, and $M_r$ is the mass within radius $r$.
The hydrostatic equations are solved using the fourth-order Runge-Kutta method with adaptive stepsize control.
We use the shooting method, varying the initial central density to match the boundary condition $M(r=R_*)=M_*$, $\rho(r=R_*)=0$, and $P(r=R_*)=0$.

\begin{figure}
\epsscale{0.48}
\plotone{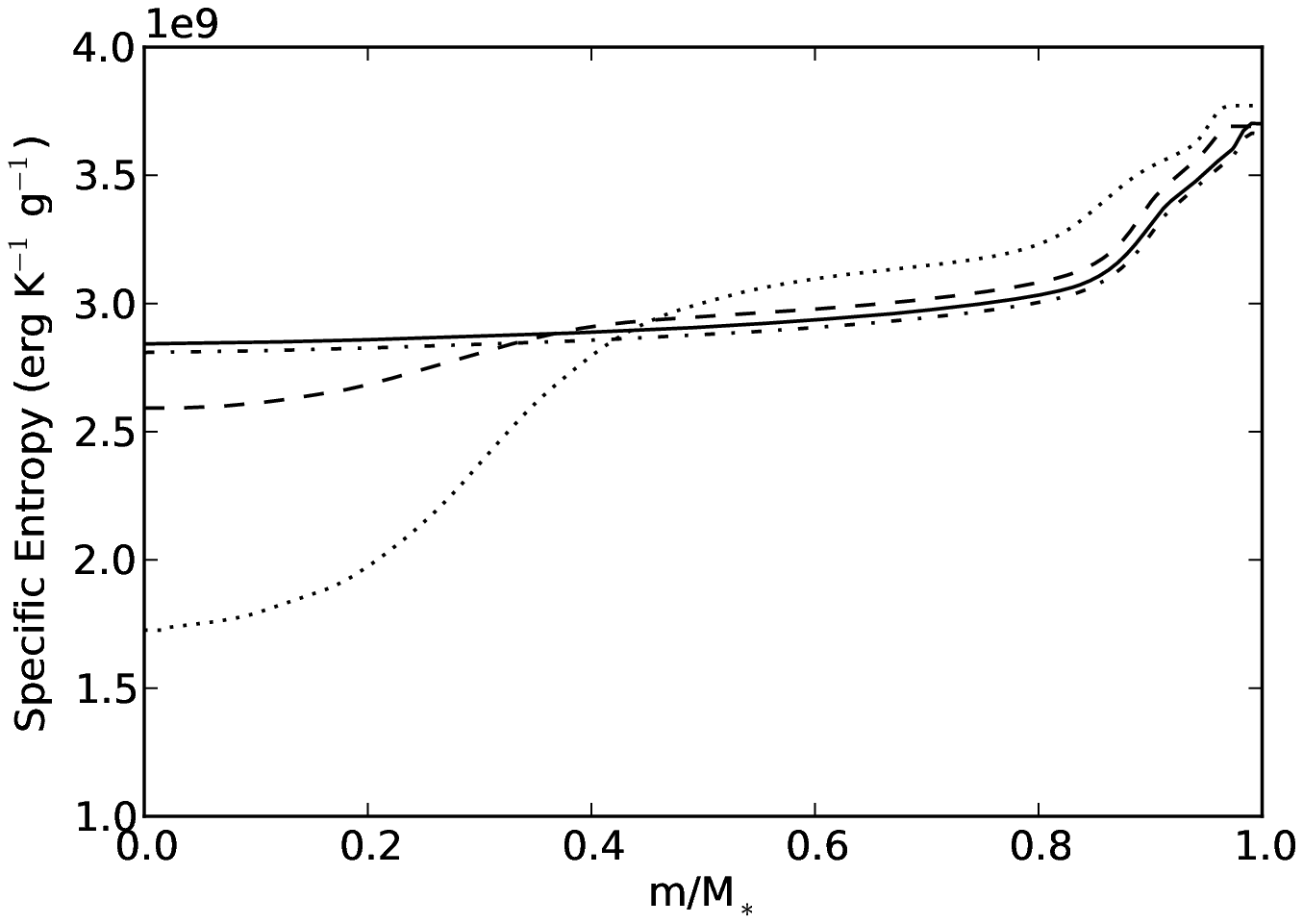}
\plotone{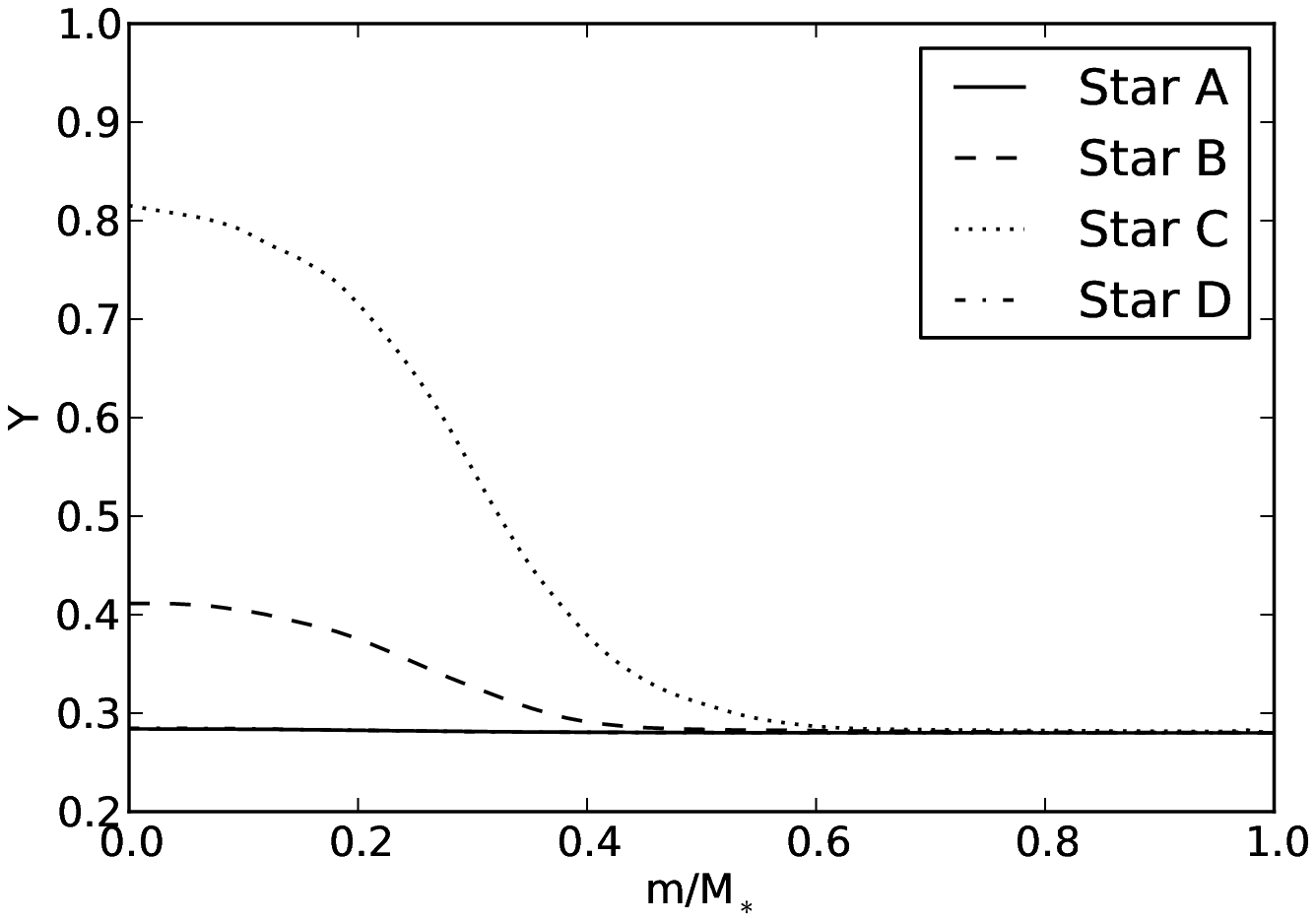}
\plotone{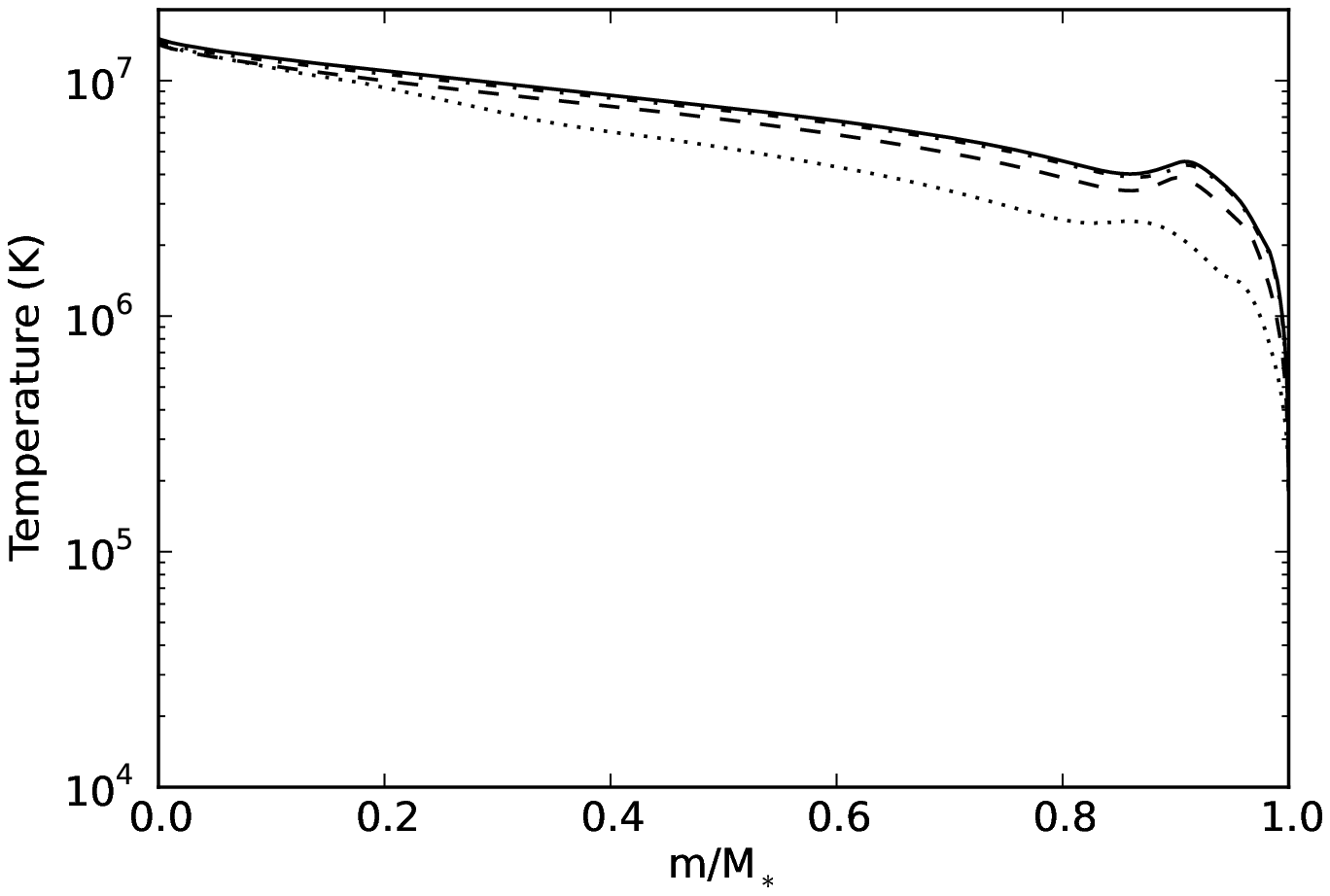}
\plotone{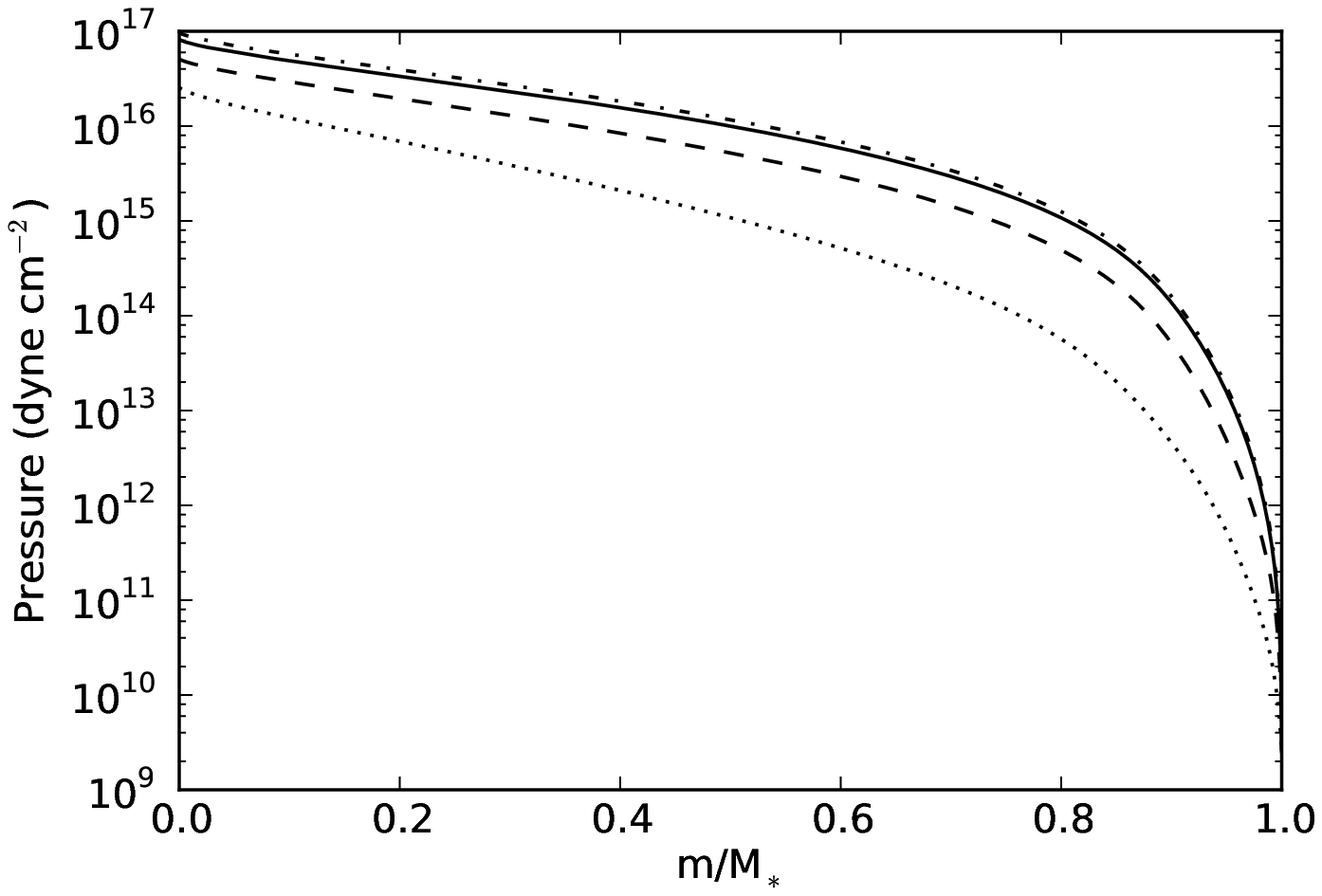}
\plotone{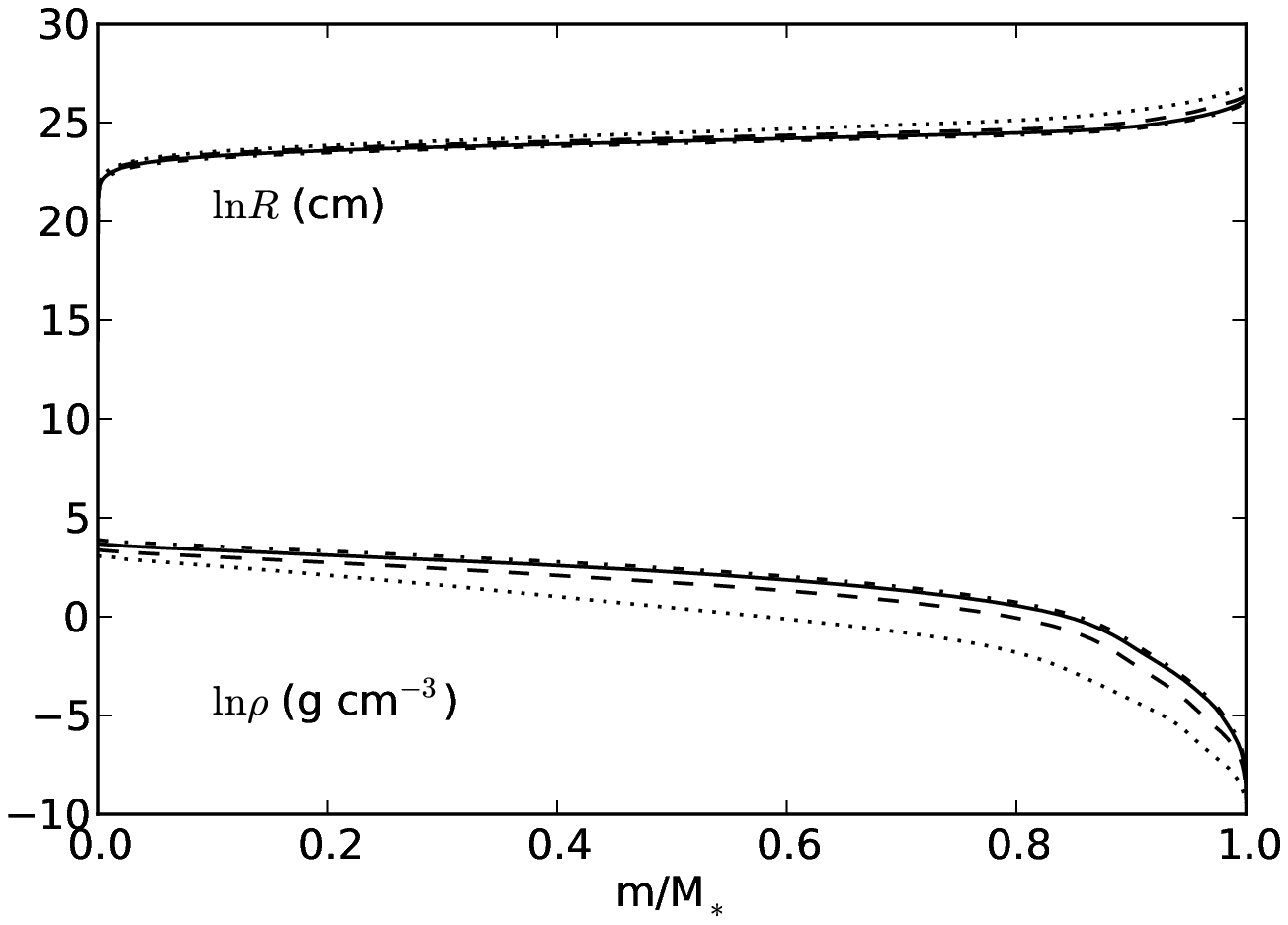}
\plotone{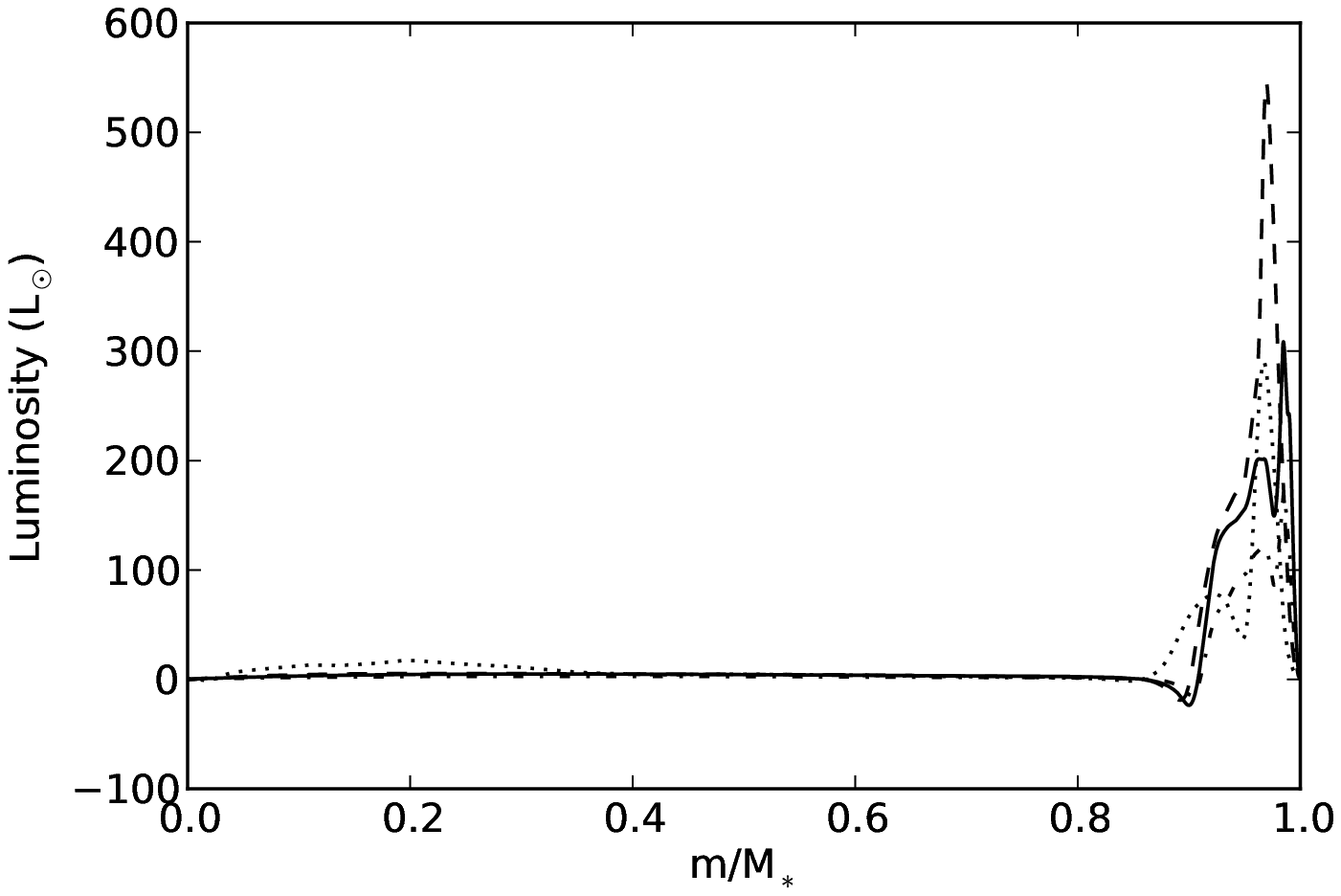}
\caption{\label{fig_profiles1} Hydrostatic solutions for entropy ($S$) helium composition (Y), density ($\rho$), temperature ($T$), pressure ($P$),
radius ($R$), and enclosed luminosity $L(m)$ for models A, B, C, and D in Table~\ref{tab_post_models}.}
\end{figure}

\begin{figure}
\epsscale{0.48}
\plotone{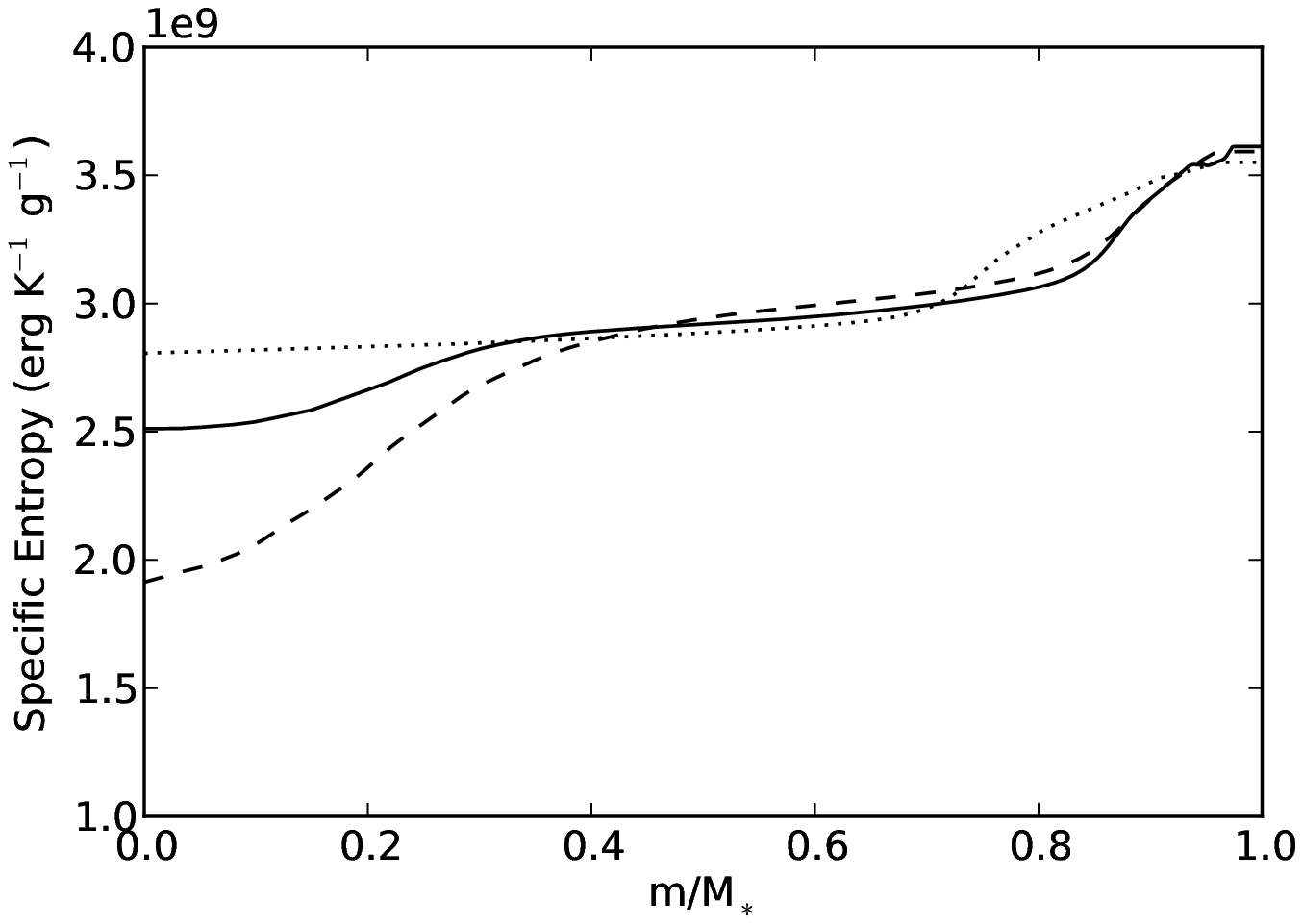}
\plotone{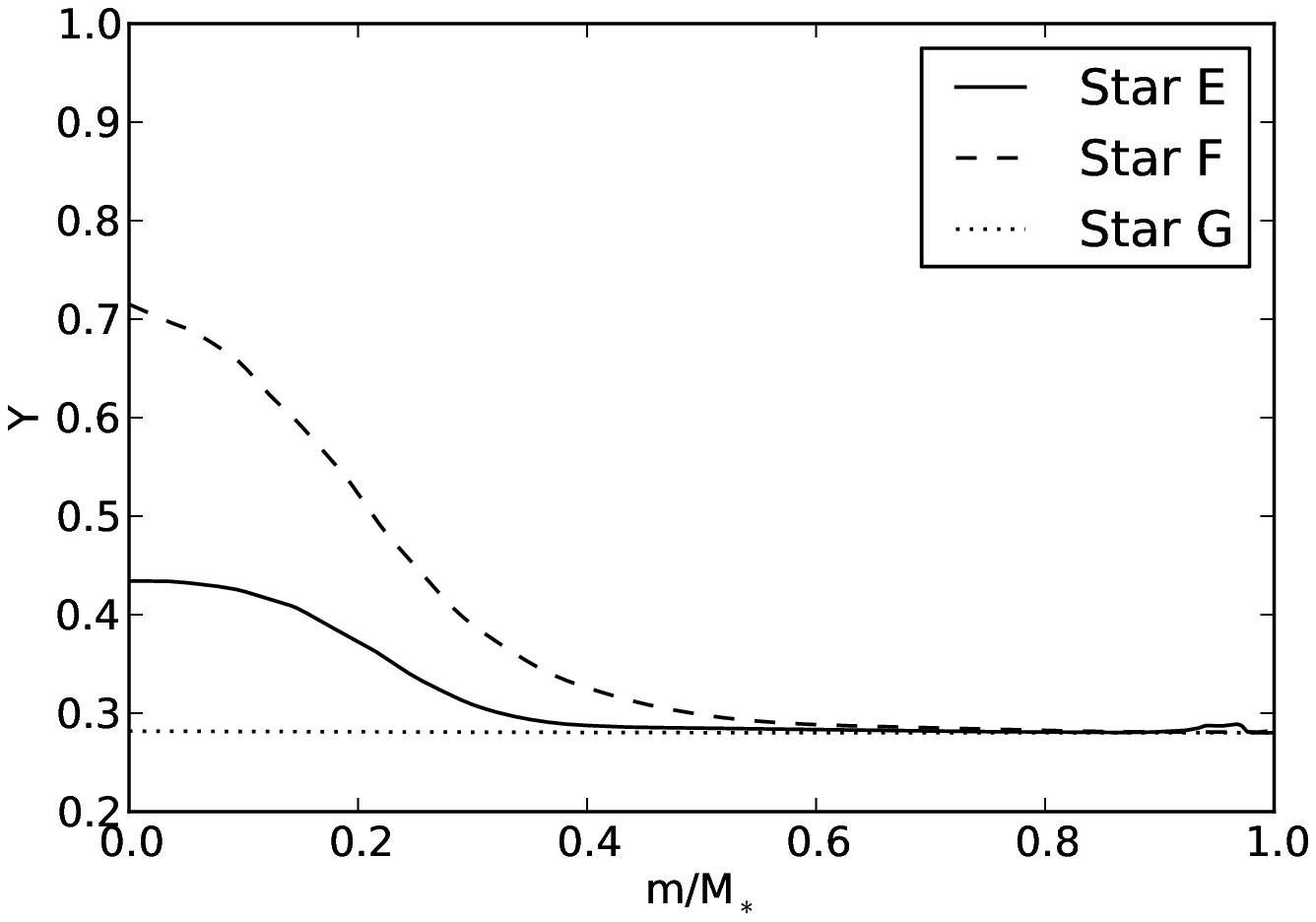}
\plotone{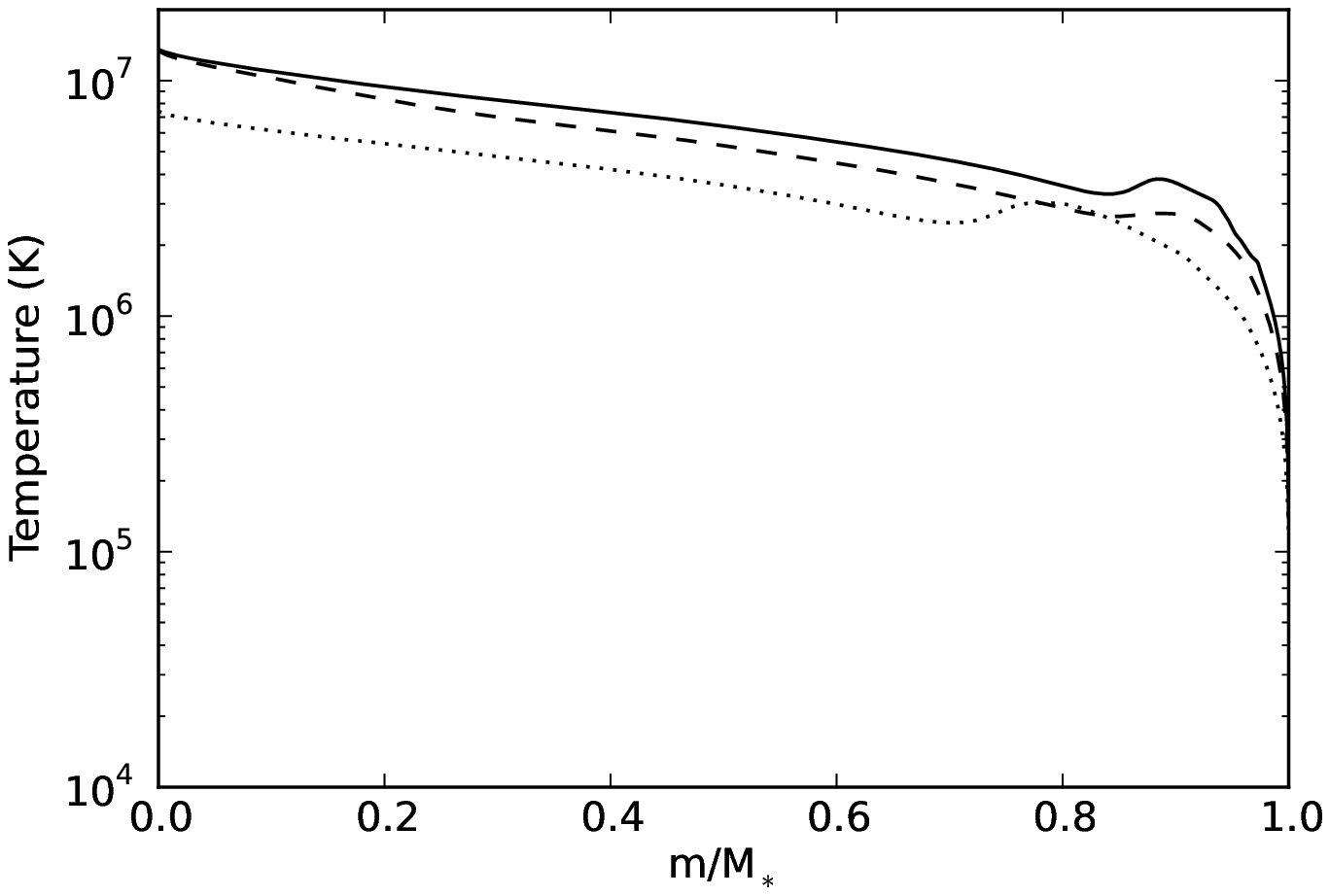}
\plotone{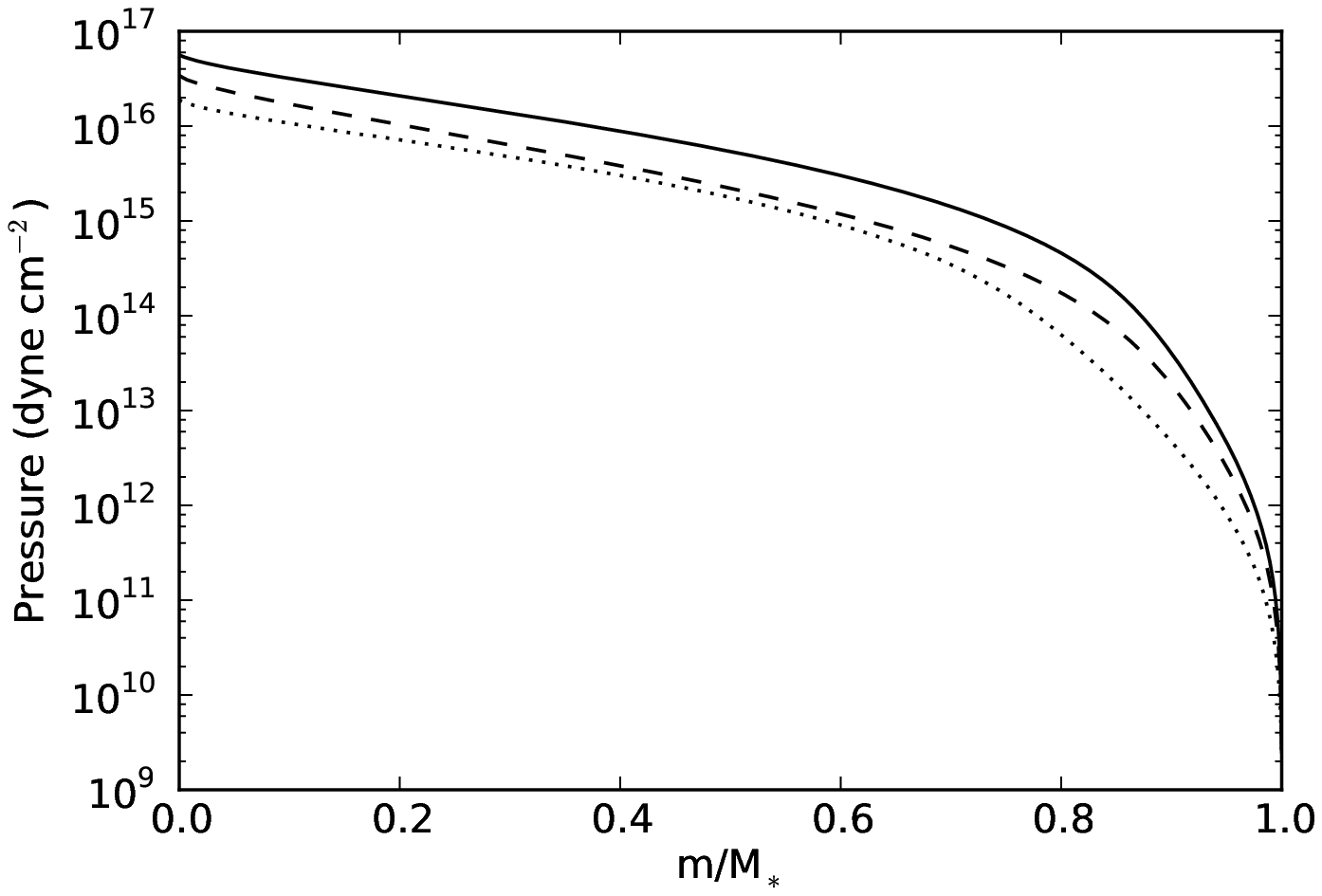}
\plotone{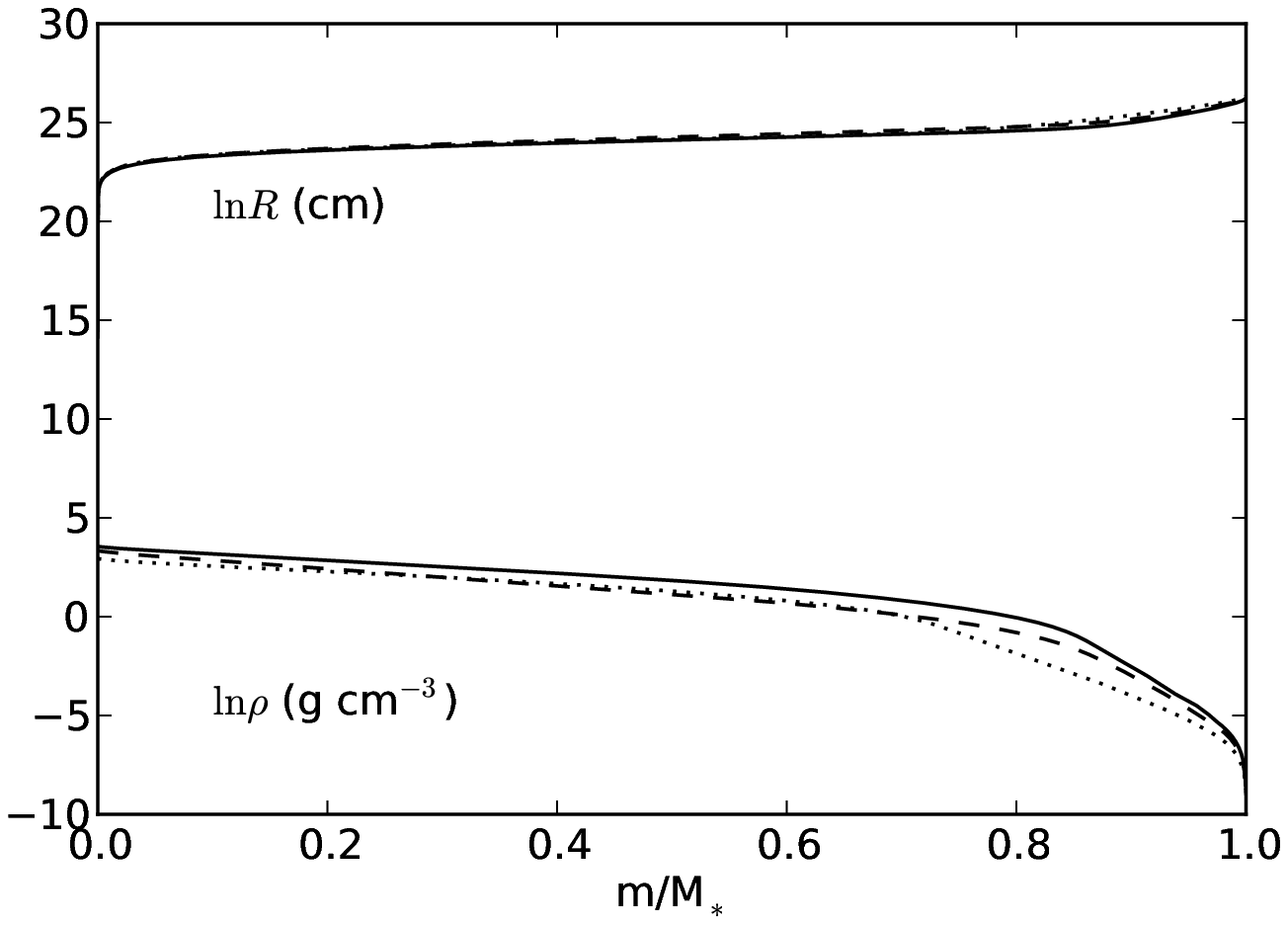}
\plotone{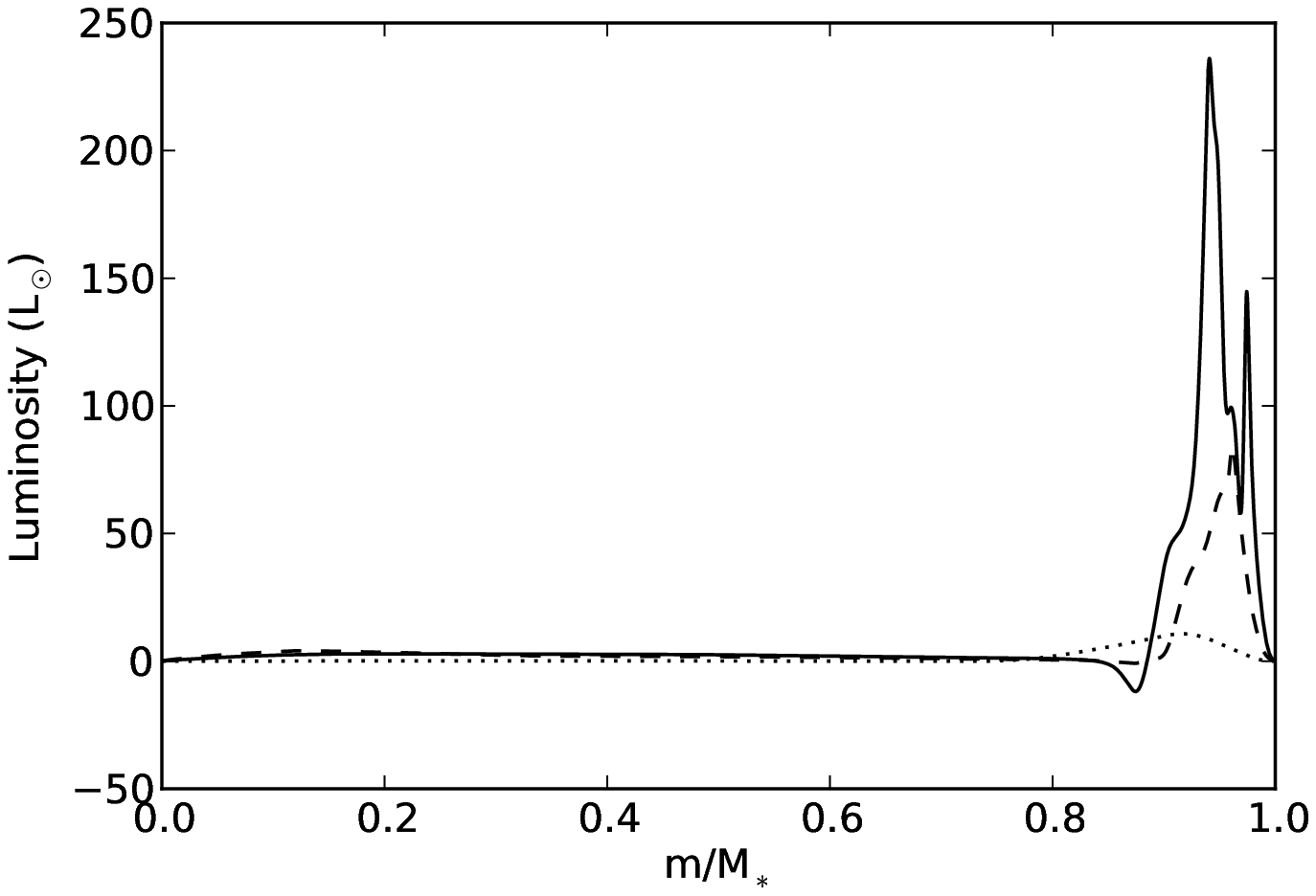}
\caption{\label{fig_profiles2} Similar to Figure~\ref{fig_profiles1} but for models E, F, and G.}
\end{figure}

The reconstructed mass and radius of the post-impact remnant stars are shown in Figure~\ref{fig_rm}. 
The upper left and right panels in Figure~\ref{fig_profiles1} and Figure~\ref{fig_profiles2} show the averaged 1D entropy and helium
composition profiles from the FLASH output.
Models B, C, E, and F have larger radii and higher central helium abundances (upper right panel), 
since their delay time and initial orbital periods are longer in HKN.
We flatten the entropy profiles in the outermost region ($ 0.95-0.99 < m/M_* < 1$) to avoid negative entropy gradients.
The sensitivity of our results to these flattened entropy profiles have been tested by varying the amount of the flattened entropy and
 using a smoothly increasing profile instead of a flat one. 
We find that the outermost region of the entropy profile does not significantly affect the hydrostatic solution and post-impact evolution. 

Another necessary variable is the initial luminosity profile, $L(m)$. 
If there is no convection, the luminosity profile can be estimated using the radiative temperature gradient expression,
\begin{equation}
L(m)= - \frac{(4\pi r^2)^2 ac}{3 \kappa}\frac{dT^4}{dm}\ ,
\end{equation}
where $\kappa$ is the opacity and $c$ is the speed of light. 
The opacity can be calculated using the {\tt kap} module in MESA for a given density, temperature, and composition.
This estimate is reasonable only at the beginning, since the modified entropy gradient is positive everywhere. 
The energy deposited by SN ejecta heating 
causes a small temperature bump at $m \sim 0.95 M_*$ (see middle panels in Figure~\ref{fig_profiles1} and Figure~\ref{fig_profiles2}), 
leading to the luminous region in the luminosity profiles in Figure~~\ref{fig_profiles1} and Figure~\ref{fig_profiles2}.


\section{EVOLUTION OF THE POST-IMPACT REMNANT STAR \label{sec_evolution}}
In this section, the post-impact evolution of remnant stars using MESA is qualitatively described 
and compared with Tycho G.  The effect of post-impact evolution with a different explosion energy 
is also studied to provide an understanding of the resultant evolution for a non-W7 explosion. 
In addition, the angular momentum distribution of post-impact remnant stars is examined by using the specific angular 
momentum from FLASH and the hydrostatic models in MESA, addressing the rotation problem in Tycho G. 

\subsection{Post-Impact Evolution}

\begin{table}
\caption{Post-impact remnant star after 440 years\label{tab_440yrs}}
\begin{tabular}{cccccc}
\\
\tableline
Model & M$_{\rm SN}$ ($M_\odot$) \tablenotemark{a}  & $R_{\rm SN}$ ($R_\odot$) & $L_{\rm SN}$ ($L_\odot$) & $T_{\rm eff,SN}$ (K) & $v_{\rm Rotation}$ (km~s$^{-1}$) \\
\tableline 
A & 1.64 & 7.18 & 202.7 & 8,135 & $10.5 \pm 1.3$\\
B & 1.65 & 11.9 & 279.0 & 6,832 &$8.3 \pm 0.6$ \\
C & 1.56 & 10.5 & 81.4 & 5,356 & $2.0 \pm 2.9$\\
D & 1.43 & 6.57 & 111.4& 7,319 & $9.1 \pm 1.3$\\
E & 1.44 & 4.56 & 20.2 & 5,737 &$26.7 \pm 4.4$ \\
F & 1.30 & 4.35 & 17.0 & 5,623 &$17.3 \pm 2.2$\\
G & 0.93 & 4.37 & 13.4 & 5,287 &$18.9 \pm 2.5$\\
GH09 \tablenotemark{\dagger}& 1.0 & $1.4-2.8$ & $1.9 -7.6 $ & $5900 \pm 100$ & ---\\
GH09 \tablenotemark{\dagger} & 1.4 & $1.6-3.3$ & $3.0 - 11.8 $ & $5900 \pm 100$ & ---\\
\end{tabular}
\tablenotetext{a}{The mass ($M_{\rm SN}$), radius ($R_{\rm SN}$), 
luminosity ($L_{\rm SN}$), effective temperature ($T_{\rm eff,SN}$), and surface rotational speed ($v_{\rm Rotation}$) of post-impact remnant stars at
440~yr after SN explosion.}
\tablenotetext{\dagger}{The radius and luminosity are estimated using the surface gravity value $\log{g} = 3.85 \pm 0.3$ in \cite{2009ApJ...691....1G},
assuming the mass is $1 M_\odot$ or $1.4 M_\odot$.} 
\end{table}

The hydrostatic solutions described in the previous section are used as initial conditions for the models used in MESA.
The initial timestep in MESA is chosen to be between $10^{-4}$ and $10^{-1}$ years, depending on the particular model. 
The compositions in the FLASH simulations are expanded to the extended network in MESA with 25 isotopes.
The percentages of compositions other than $^1$H and $^4$He are scaled to solar abundances, but with the same total metallicity.  
When starting a MESA simulation the model is iterated until convergence to ensure 
that both the EOS and the energy generation rate are consistent. 
During the first few steps, the evolution is transient, but eventually evolves to a parameter-independent model within a year.   
Thus, only simulation results after one year are considered in this study. 

\begin{figure}
\epsscale{0.55}
\plotone{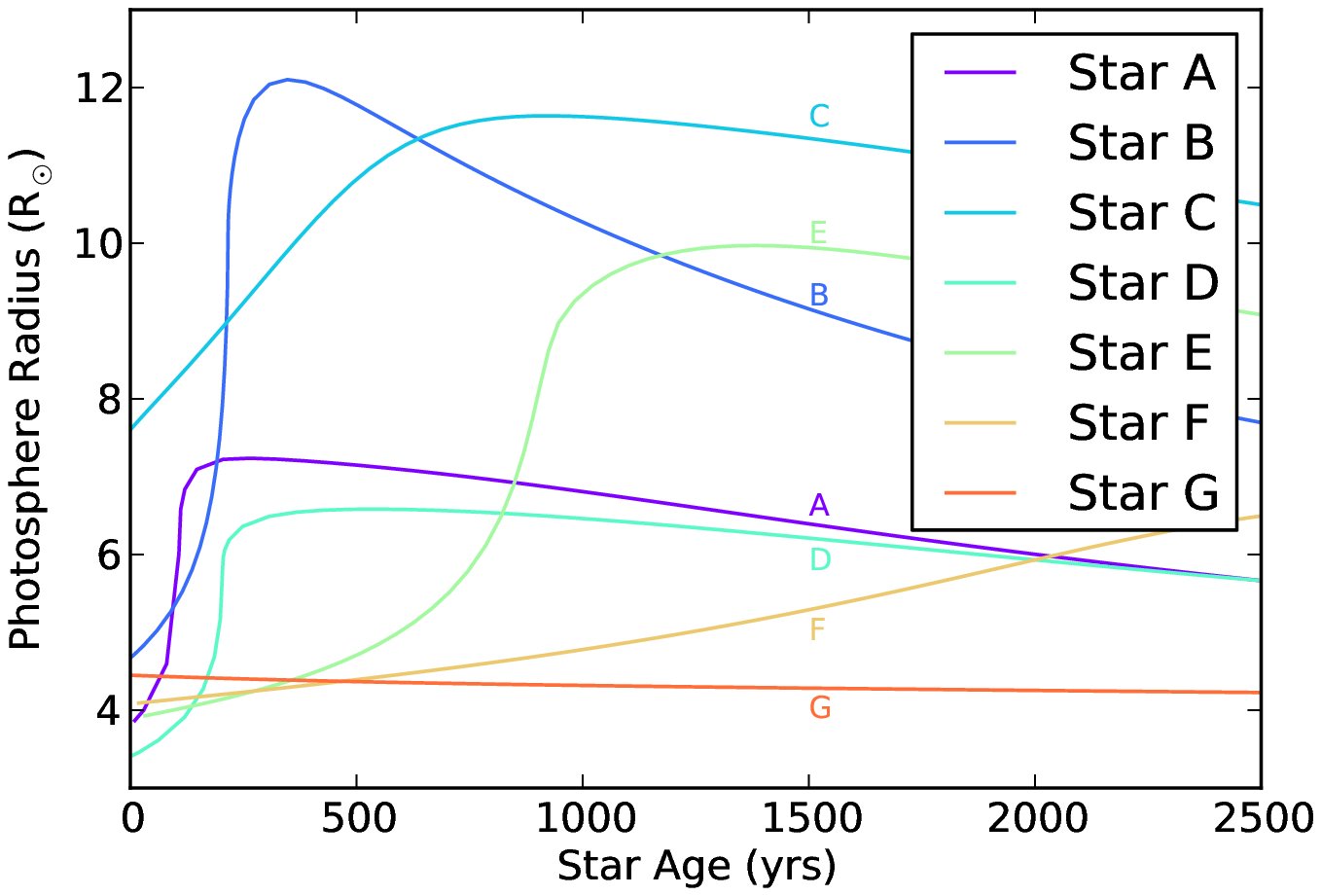}
\plotone{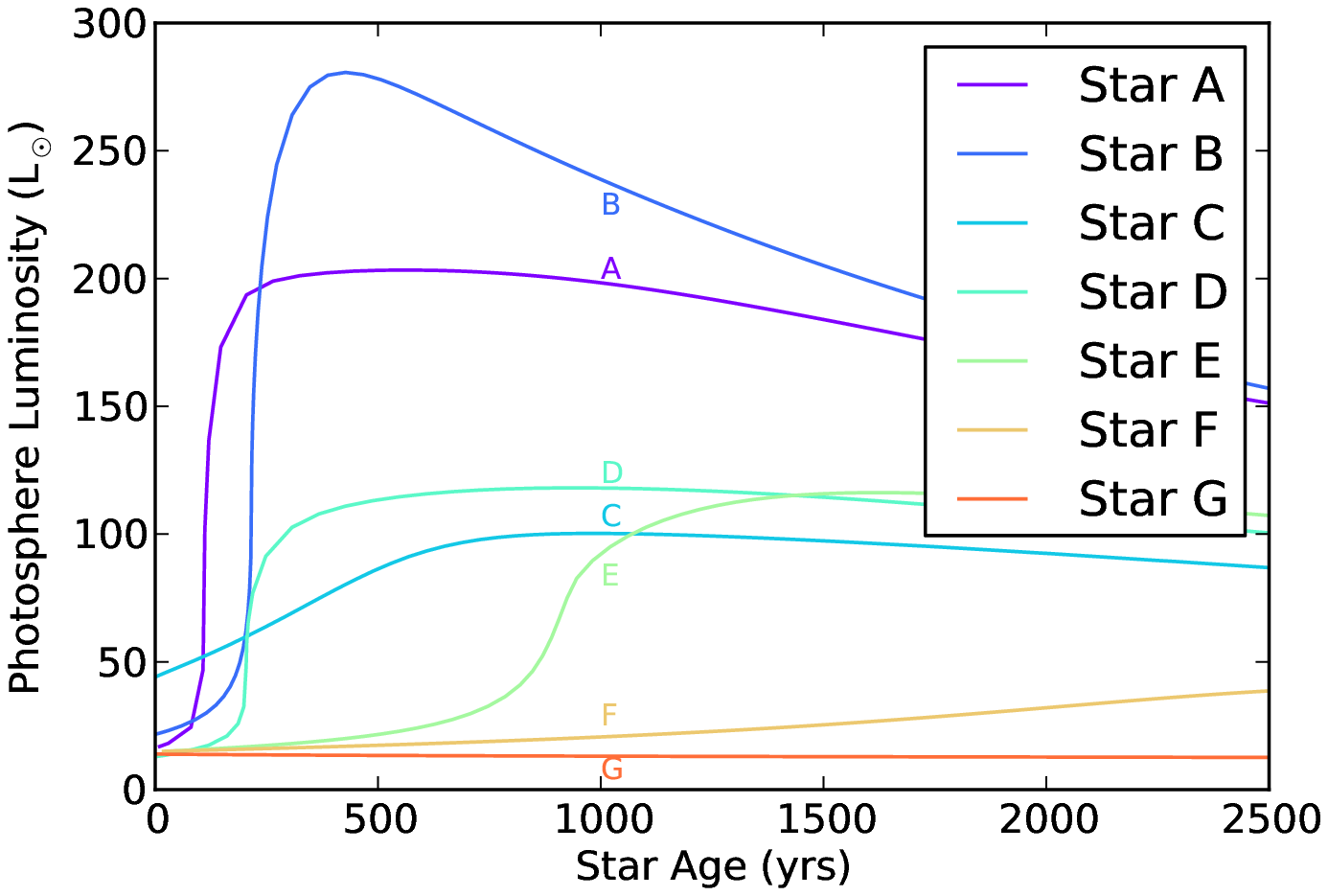}
\plotone{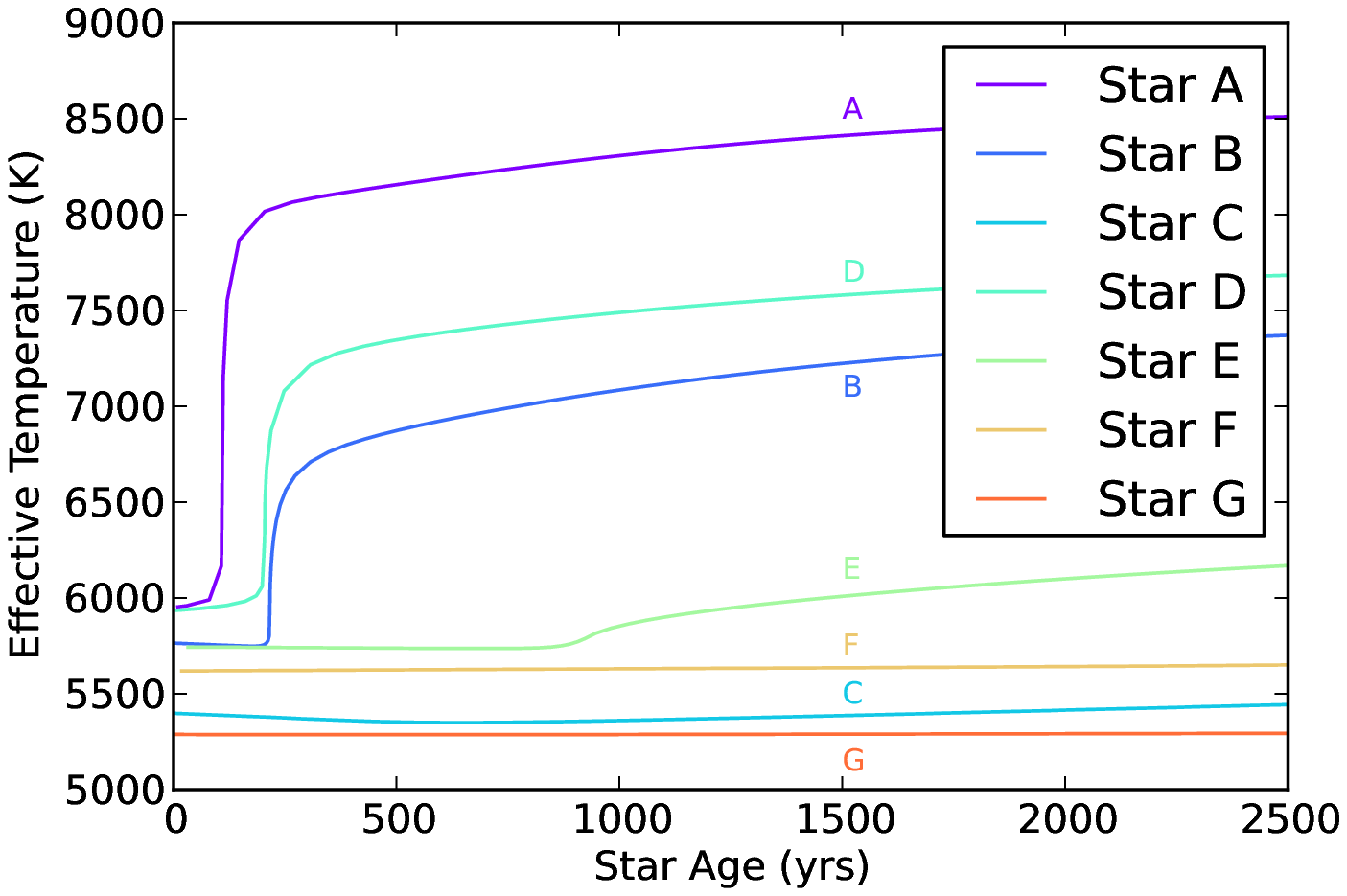}
\caption{\label{fig_evolve}
Evolution of the photosphere radius, luminosity, and effective temperature as functions of time. 
Each line shows the evolution of a post-impact companion star in Table~\ref{tab_post_models}.}
\end{figure}

Figure~\ref{fig_evolve} shows the evolution of the photospheric radius, luminosity, and effective temperature as functions of time. 
The post-impact remnant stars rapidly expand on a timescale of $\sim 10^2$ years for models A, B, D, and E, and 
$\sim 10^3$ years for models C and F.  
Note that there is no significant change in model G within $10^4$ years, 
because the thermal timescales in model G are much longer than other stars.

Here we describe the detailed evolution of model A as an example.
Due to the high opacity in the outermost region of the envelope, 
the photospheric luminosity is $\sim 4$~$L_\odot$, 
but the luminosity inside the envelope is much higher than the surface luminosity (Figure~\ref{fig_profiles1}). 
Therefore, the strong radiation at $m \sim 0.9 M_*$ expands the outermost $\lesssim 1\%$ of mass.     
In addition, the luminosity profile of the outermost $10\%$ of mass flattens due to radiative diffusion. 
This radiative diffusion timescale is characterized by the local thermal timescale \citep{1969ApJ...156..549H},
\begin{equation}
\tau_{\rm th}(r) = \frac{3}{64 \pi \sigma_{\rm SB}}\left[ \int_{r}^{R_*} \left( \frac{\kappa C_p}{T^3}\right)^{1/2} \rho dr\right]^2\ ,
\end{equation}
where $\sigma_{\rm SB}$ is the Stefan-Boltzmann constant, and $C_p$ is the specific heat capacity.
Although the global thermal timescale, $t_{\rm th} = GM^2/2RL$, is of the order of $10^6$ years, 
the local thermal timescale in the envelope region is only $\sim 10^2-10^3$ years.

After $\sim 200$ years, the deposited energy has been radiated away and model A starts to contract by releasing gravitational energy.
The luminosity decreases, but the effective temperature increases.  The star will eventually return to a stage similar to 
the ZAMS on a global thermal timescale.  As an SN~Ia remnant may not be recognizable in millions of years, the simulation 
is terminated at $10^4$ years.  

\begin{figure}
\epsscale{0.9}
\plotone{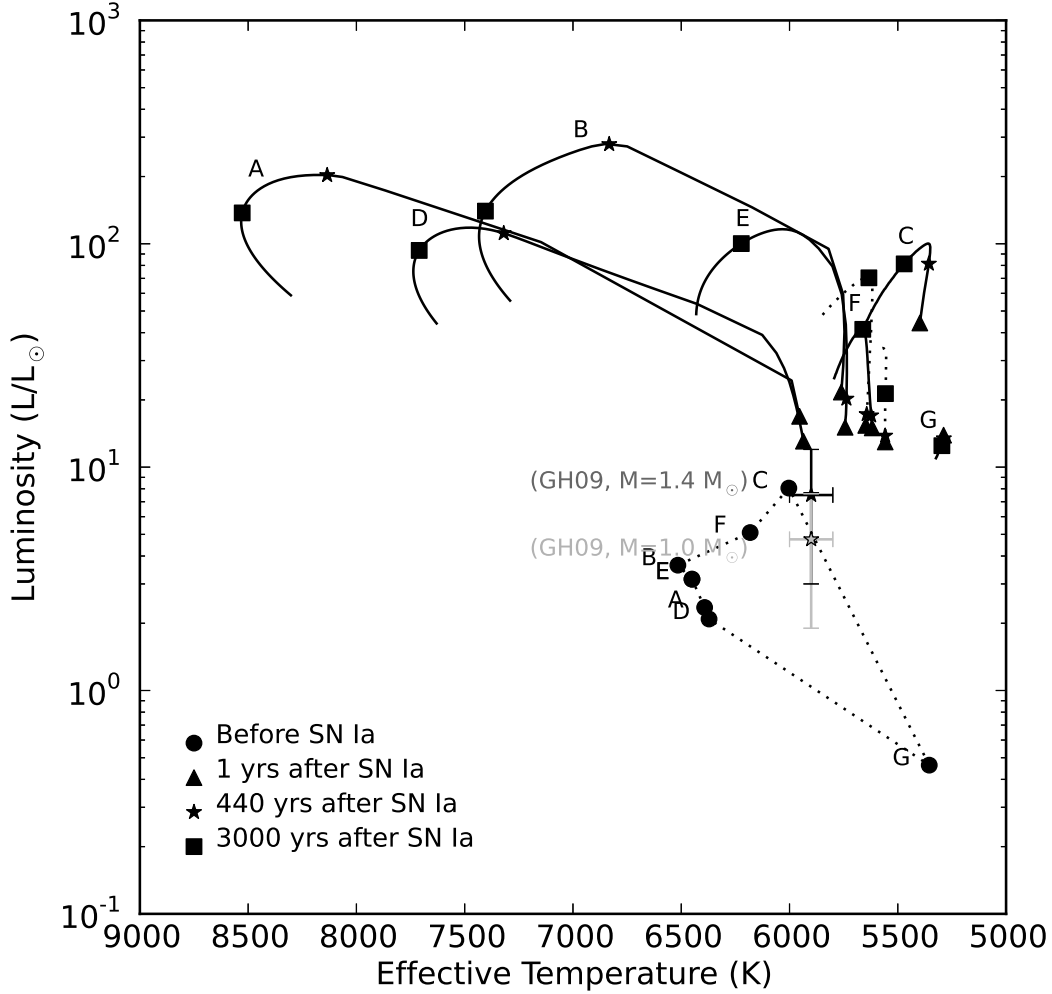}
\caption{\label{fig_HR} 
The evolutionary tracks in the HR diagram for different post-impact companion models. 
Each line shows an evolutionary track of a companion model in Table~\ref{tab_post_models} over an interval of $10^4$~years. 
The filled circles indicate the condition of stars just before the SN~Ia explosion; 
filled triangles present the condition of stars at $\sim 1$ year after the SN impact;
star symbols show the conditions at $440$ years after the SN impact, 
which is the same age as Tycho's SN;
filled squares present the conditions at $3,000$ years after the SN impact.
The star symbols with error bars show the observed luminosity and effective temperature of Tycho G as measured by 
\cite{2009ApJ...691....1G} (GH09) in Table~\ref{tab_440yrs}.}
\end{figure}

The evolutionary tracks of post-impact remnant stars in the Hertzsprung-Russell (H-R) diagram are plotted in Figure~\ref{fig_HR}.  
In general, the remnant stars become brighter but cooler after the impact.    
By comparing Figure~\ref{fig_HR} with Figure~\ref{fig_rm},
we find that there is a trend for more massive stars coupled with smaller radius (shorter orbital period) to evolve faster in the HR diagram. 
In addition, stars with smaller radii have higher effective temperatures after the SN impact.
This can be understood given the fact that stars with shorter orbital periods are closer to the exploding WD, leading to more violent
SN impacts and greater energy deposition, but with less unbound mass due to their more compact states. 

The Tycho G star, based on observations by 
\cite{2009ApJ...691....1G}, is also plotted in Figure~\ref{fig_HR}.
\cite{2004Natur.431.1069R} determined Tycho G's effective temperature, $T_{\rm eff} = 5750 \pm 250$~K, and surface gravity,
$\log (g/ {\rm cm}$ {\rm s}$^{-2})= 3.5 \pm 0.5$, by fitting spectral lines. 
Subsequently, \cite{2009ApJ...691....1G} updated the observed values to
$T_{\rm eff}=5900 \pm 100$~K and $\log (g/ {\rm cm}$ {\rm s}$^{-2})= 3.85 \pm 0.30$ based on fits to iron lines.
\cite{2009ApJ...691....1G} estimated the radius of Tycho G using the surface gravity, 
assuming the mass of Tycho G to be $1M_\odot$. 
They estimated the bolometric luminosity $L_*$ to lie in the range $1.9 < L_*/L_\odot < 7.6$. 
However, this luminosity could be underestimated if Tycho G is more massive.
Since $\log L_* \propto 2\log R + 4\log T $ and $\log g \propto \log M - 2\log R$, 
$L_*$ could be increased to $3.0 < L_*/L_\odot < 11.8 $ if the mass of Tycho G were instead $1.4 M_\odot$.  
Our post-impact remnant stars have masses ranging from $1.3M_\odot$ to $1.65 M_\odot$.
A mass $M_*=1.4M_\odot$ instead of $1M_\odot$ is, therefore, assumed for Tycho G.  
As a result, the placement of Tycho G in the HR diagram is now closer to our models A, B, D, E and F
immediately after the SN impact (triangle symbols in Figure~\ref{fig_HR}). 
However, Tycho's SN exploded 440 years ago, 
and models A, B and D would have evolved to a hotter and more luminous state in that time (star symbols in Figure~\ref{fig_HR}), 
suggesting that model E among our progenitor systems is the least discrepant with Tycho G.
Our model E has a consistent effective temperature, but the luminosity (radius) is twice brighter (larger) than the observed value in \cite{2009ApJ...691....1G}.
Model F has a luminosity similar to that of model E, but the effective temperature is a few hundred degrees lower than Tycho G. 
A less massive or smaller model than model E may better match the observation by \cite{2009ApJ...691....1G}.
Table~\ref{tab_440yrs} summarizes the stellar properties for all the progenitor models at 440~yr after the SN explosion\footnote{
Two recent observations of SN~1006 by \cite{2012Natur.489..533G} and \cite{2012arXiv1207.4481K}
suggest that there are no surviving evolved companions in the central region of SN~1006.  
We have noted that the star B90474 ($\log g = 3.05 \pm 0.12~({\rm cm~s}^{-2}$), $T_{\rm eff}= 5051 \pm 38~({\rm K})$) 
and B14707 ($\log g = 3.36 \pm 0.15~({\rm cm~s}^{-2}$), $T_{\rm eff}= 5065 \pm 47~({\rm K})$) 
have similar surface gravity as our 
Model G ($\log g = 3.14~({\rm cm~s}^{-2}$), $T_{\rm eff}= 5288~({\rm K})$) but a lower effective temperature.
However, the star B14707 lies much closer than the SNR and the distance uncertainity of star B90474 is large. 
}.

\subsection{Dependence on the SN Ejecta Energy}
Although Tycho's SN has previously been suggested to be a subluminous SN~Ia  \citep{1993ApJ...413...67V}, or a overluminous SN~Ia \citep{2004ApJ...612..357R},  
the most recent observation using the light echo suggests that Tycho's SN is actually a normal SN~Ia \citep{2008Natur.456..617K}. 
In our hydrodynamics simulations, the W7 model ($E_{\rm SN} = 1.23 \times 10^{51}$~erg) is assumed to describe the SN explosion.
However, the exact explosion energy could be different from case to case. 
Therefore, two additional simulations for models D and E, which are the remnant stars closest to Tycho G, are performed with double the explosion energy. 
We specify that this additional energy is in a thermal form. 

\cite{2008A&A...489..943P} have shown that the unbound mass of the companion star is linearly proportional to the SN kinetic energy. 
In our hydrodynamics simulations, a similar trend is observed, but it is not linear, 
since there are essentially two mechanisms that unbind the mass: ablation and stripping (see PRT for detailed description). 
The resulting mass of the post-impact remnant star for model D (model E) is $M_{\rm W7} =$ 1.43 $M_\odot$ (1.44 $M_\odot$) for the W7 case and $M_{\rm 2X} =$ 1.23 $M_\odot$ (1.37 $M_\odot$) for 
the case with double the explosion energy.

\begin{figure}
\plotone{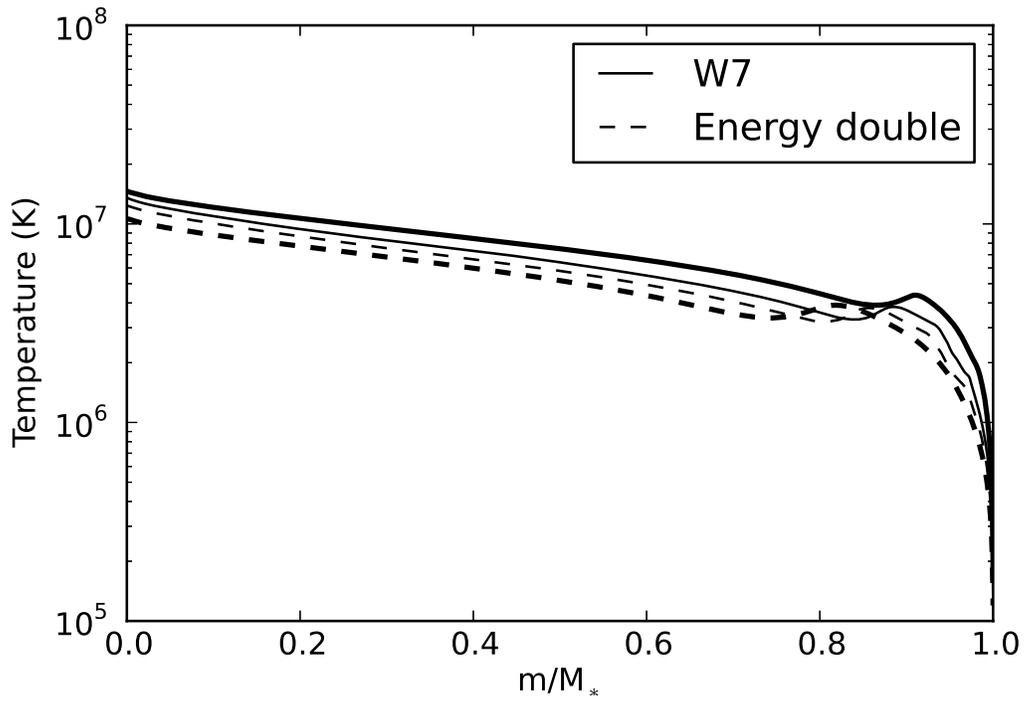}
\caption{\label{fig_energy_t}
Temperature vs. mass profiles of the hydrostatic solutions for model D (thick lines) and model E (thin lines) with different SN~Ia explosion energies. 
The solid lines indicate the temperature profiles with explosion energy using the W7 model ($E_{\rm SN} = 1.23 \times 10^{51}$~erg). 
The dashed lines show the temperature profiles with double the explosion energy.}
\end{figure}

Figure~\ref{fig_energy_t} shows the temperature profiles of these four explosion cases after the SN impact.
The heating is stronger in the higher explosion energy case, resulting in a lower remnant mass and causing a lower central temperature. 
The temperature bump in the higher explosion energy case shifts inward because the heating penetrates to deeper mass layers 
with a stronger impact.  Therefore, the radiative diffusion time for the propagation of energy to the surface becomes longer. 

\begin{figure}
\epsscale{0.55}
\plotone{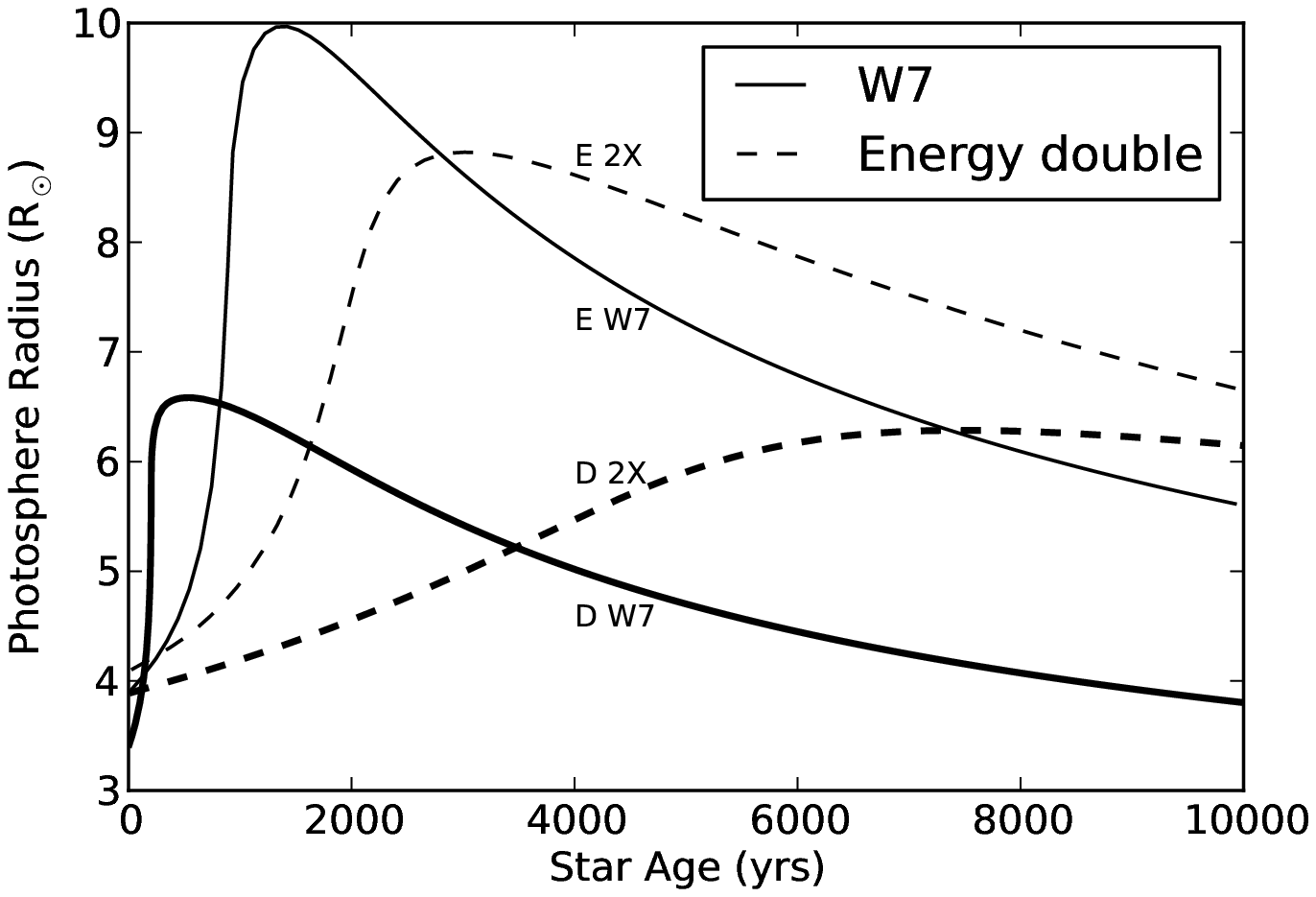}
\plotone{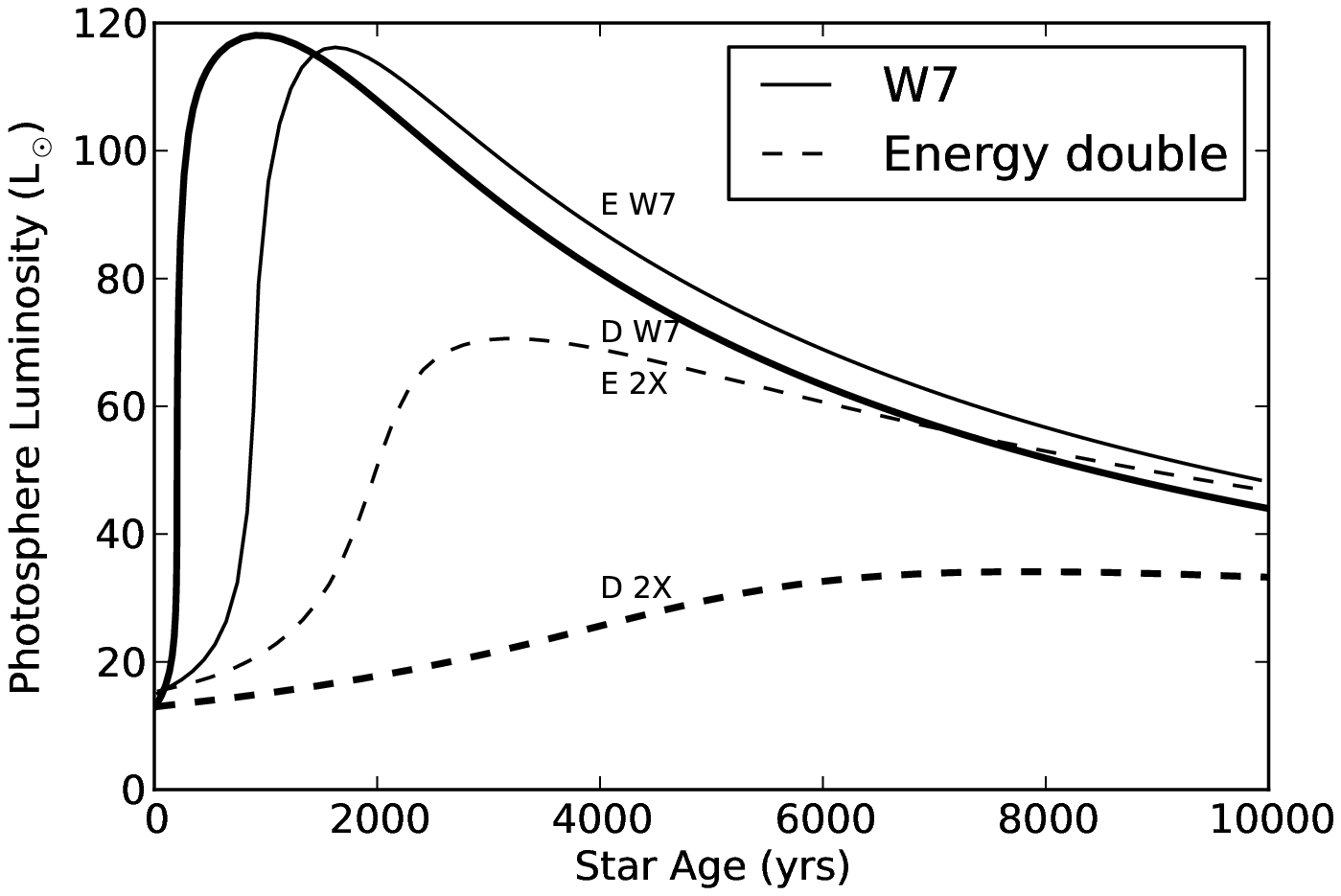}
\plotone{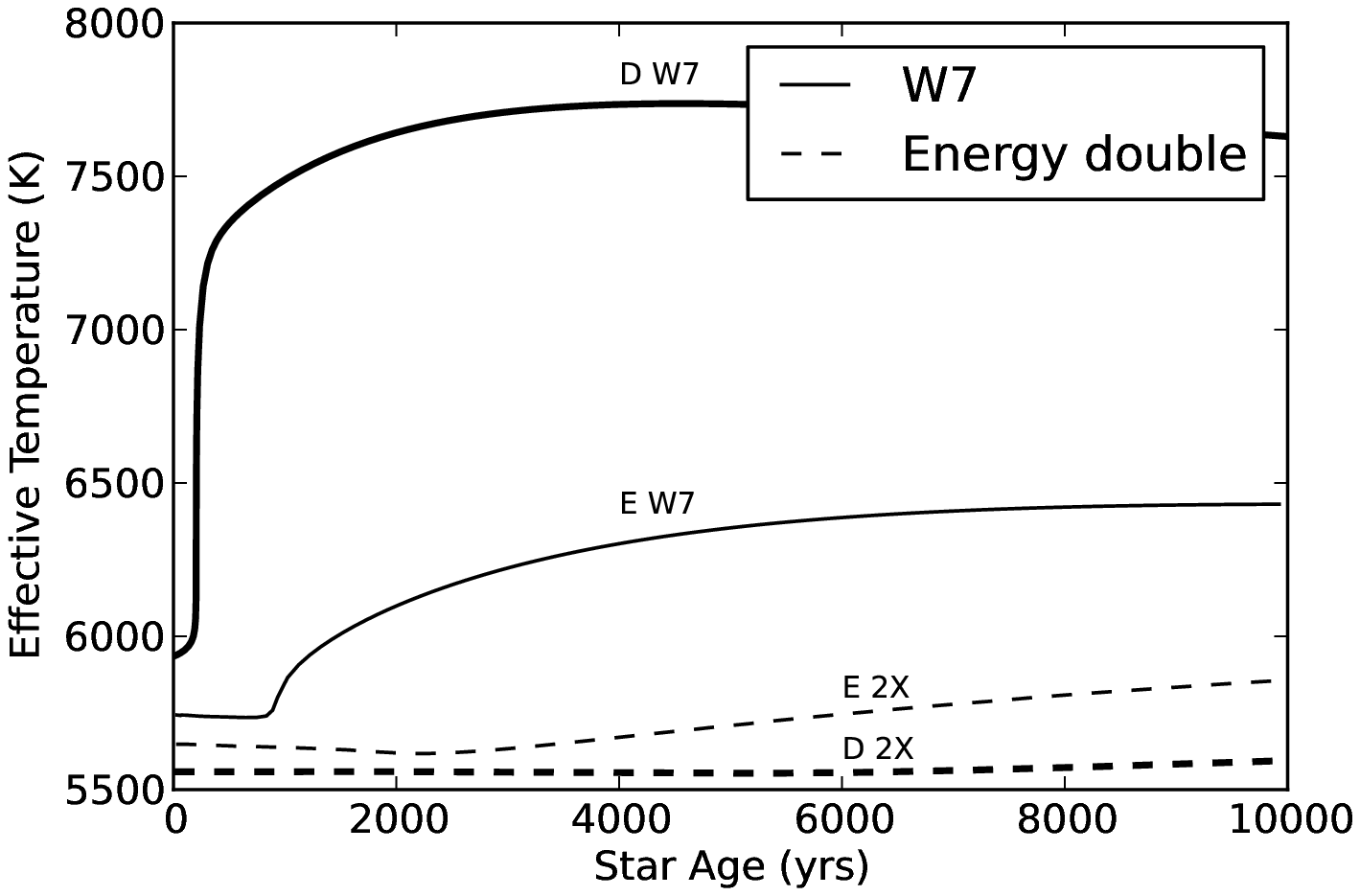}
\caption{\label{fig_energy_evolve}
Similar to Figure~\ref{fig_evolve}, but for model D (thick lines) and E (thin lines) with different SN~Ia ejecta energy.}
\end{figure}

The evolution of the photospheric radius, luminosity, and effective temperature can be seen in Figure~\ref{fig_energy_evolve}. 
The lower mass and temperature in the higher energy case cause the radius, luminosity, and effective temperature to be smaller than for the W7 case.
In general, increasing the explosion energy will lower the effective temperature and luminosity,
but the evolution timescale is also proportional to the explosion energy. For example, a dramatic difference in the 
evolution is seen for model D in which the luminosity decreases by a factor of eight
(see Figure~\ref{fig_energy_evolve}) 
at 440 years in comparison with the standard explosion energy (W7). 
This difference reflects the deeper energy deposition and longer thermal
diffusion time associated with increased explosion energy.
If we desire to match the luminosity of Tycho G by enhancing the explosion energy, 
the effective temperature becomes too low since the mass of the remnant star becomes lower.
Thus, neither an overluminous nor a subluminous explosion model for model D or E can perfectly match the observed properties of Tycho G.  

\subsection{Rotation Problem \label{sec_rotation}}

In the binary evolution scenarios in HKN, the evolution time from the onset of RLOF in the MS+WD systems to the SN~Ia phase is about $10^6$ years. 
Although the mass transfer in close binary systems may cause the companion stars to be non-corotating, 
\cite{1977A&A....57..383Z} finds the synchronization time in binary systems to be
\begin{equation}
t_{\rm sync} \sim q^{-2} (a/R)^6 \sim 10^4((1+q)/2q)^2 P^4 {\rm ~years},
\end{equation}
which is $\sim 10^3$ years in our cases.
Therefore, the companion stars should all be synchronized by tidal locking and be rapidly rotating, resulting in surface rotational 
speeds of the order of hundreds of kilometers per second.
However, the upper limit on the rotation speed of Tycho G found by \cite{2009ApJ...701.1665K} is $v \sin{i} \lesssim 7.5$~km~s$^{-1}$, 
where $i$ is the inclination angle. 

PRT indicate that the companion star (model G in Table~\ref{tab_models}) loses about half of its angular momentum while only losing about 
$20 \%$ of its mass.  With the strong assumptions that the parameter $\alpha$ is constant in the angular momentum expression, $J=\alpha M R^2 \omega$, 
and that the companion star is in a state of solid-body rotation before and after the impact, 
the rotation speed can significantly drop to about a quarter of its original rotation speed.   
However, the equilibrium status of the remnant stars and the location of the photosphere were not calculated in detail in PRT.  
Combining the post-impact angular momentum distributions in FLASH 
with the stellar structure of post-impact remnant stars in MESA, the rotation problem in the Tycho G can be addressed quantitatively.

\begin{figure} 
\epsscale{1.0}
\plotone{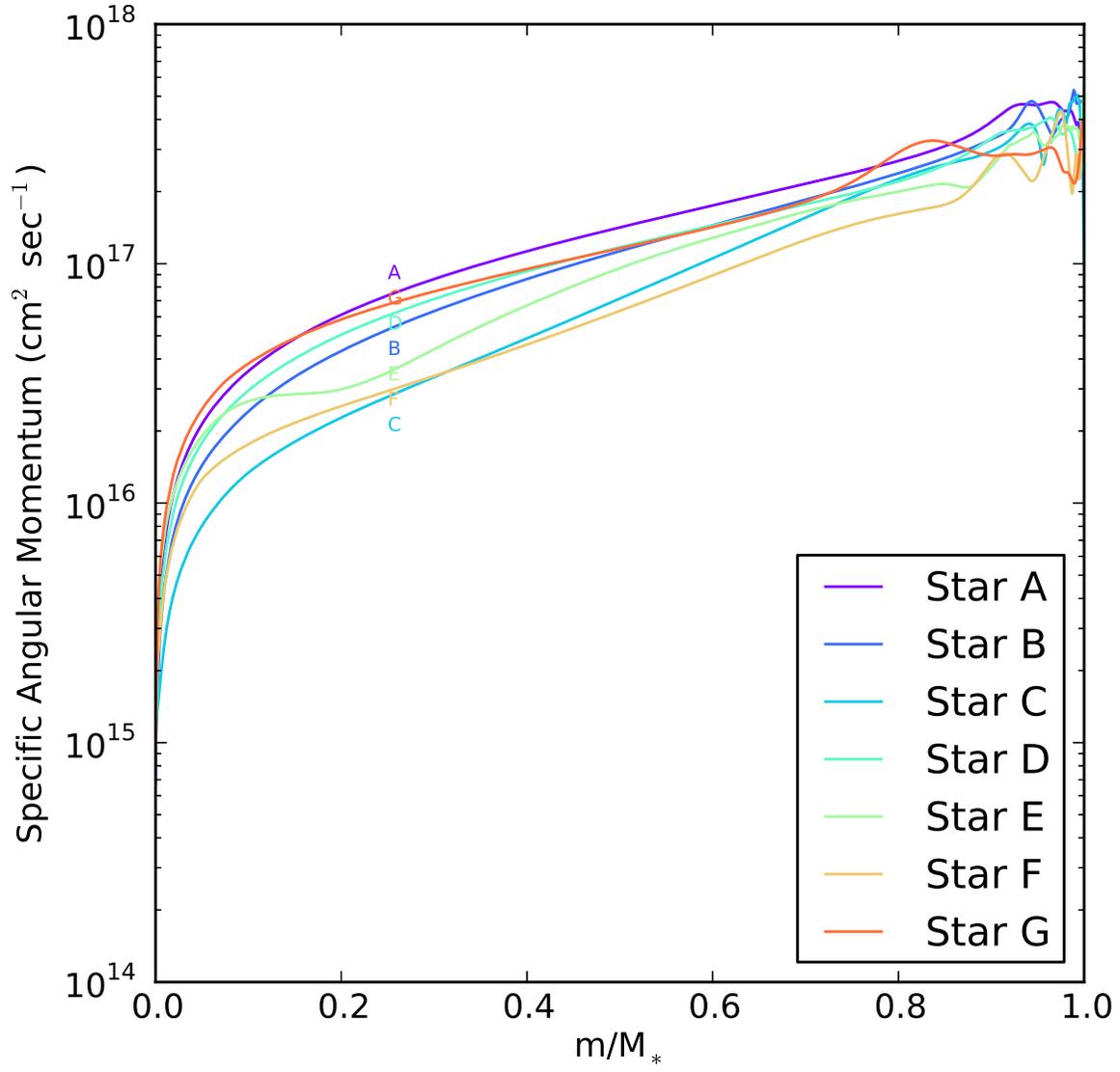}
\caption{\label{fig_sj}
Specific angular momentum as a function of the mass fraction for all companion models in Table\ref{tab_post_models}.} 
\end{figure}

Before the SN impact, the companion stars were set into uniform rotation and were given a spin with spin-to-orbit ratio of $0.95$.
After the SN impact, the companion stars are no longer in solid-body rotation. 
The angle-averaged angular velocity profiles can be calculated using
\begin{equation}
\Omega(r) = \left<\frac{v_\phi}{r \cos{\theta}} \right>_{\phi,\theta},
\end{equation}
where $\theta$ and $\phi$ are latitude (zero at the equator; positive northward) 
and longitude (positive in the direction of rotation) in spherical coordinates and in the center-of-mass frame.  
It is found that the angular velocity is insensitive to the latitude and longitude.
Thus, only the radial dependence in spherical coordinates is considered. 
If we assume the specific angular momentum, $h(m) = \frac{2}{3} r(m)^2 \Omega(m)$, to be conserved during the evolution, 
the angular velocity and surface rotational speed profiles of hydrostatic models can be calculated.
Figure~\ref{fig_sj} shows the specific angular momentum profiles of all considered models in Table~\ref{tab_post_models}.
Since the angular velocity  from the FLASH simulations has some variations in the surface region, we use the standard deviation of the specific angular momentum in the region of  $0.8 < m/M_* < 1.0$ to estimate the uncertainty of specific angular momentum.   

\begin{figure}
\epsscale{1.0}
\plotone{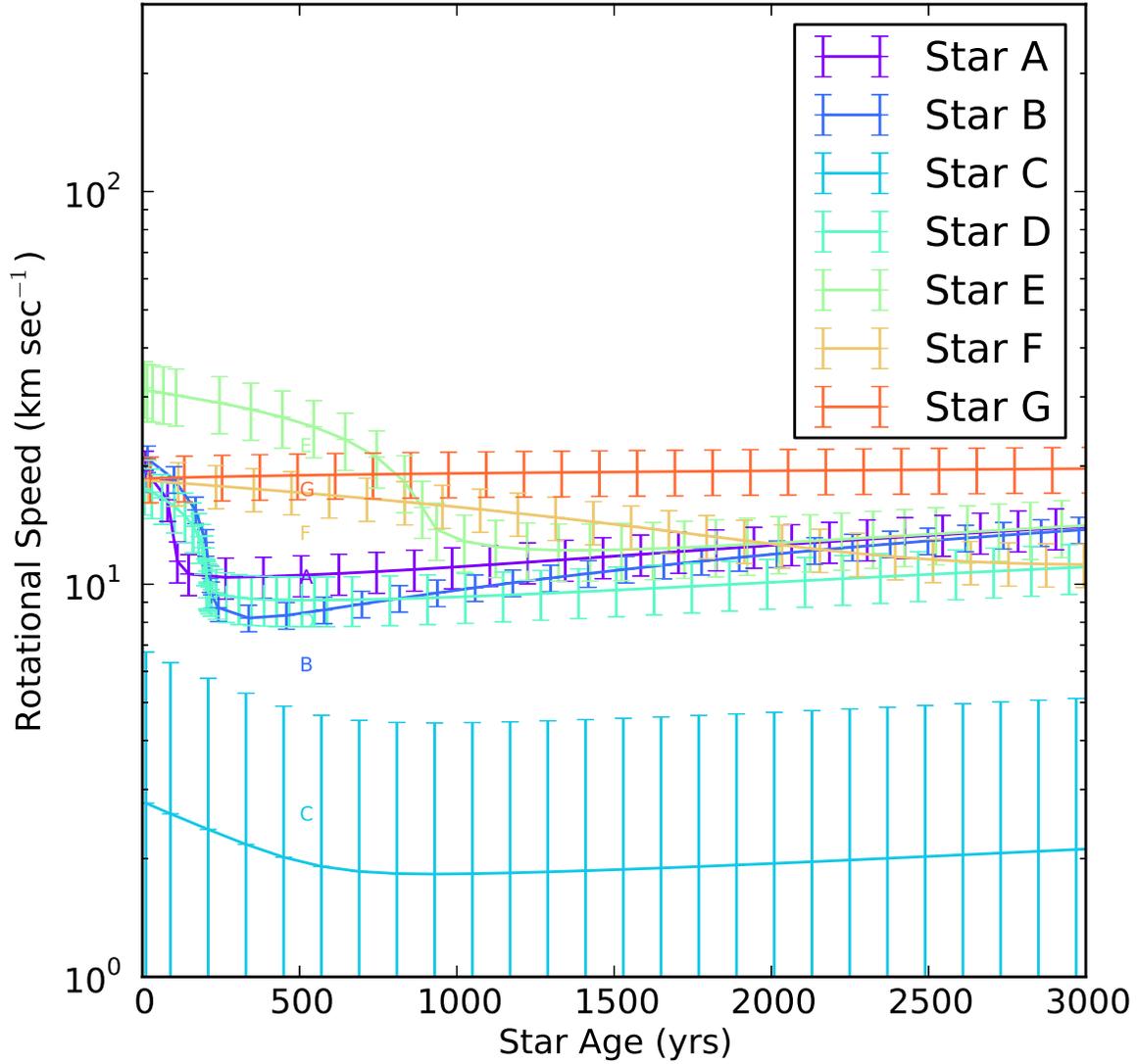}
\caption{\label{fig_vrot_evolve} 
Evolution of surface rotational speed for all the companion models in Table~\ref{tab_post_models}.
The error is estimated by the variation of specific angular momentum within $0.9< m/M_*<1.0$ in Figure~\ref{fig_sj}.}
\end{figure}

Figure~\ref{fig_vrot_evolve} shows the evolution of surface rotational speed for all the remnant star models.
The error bars show the uncertainty of surface rotational speed based on the uncertainty from the specific angular momentum.   
It is found that the rotational speed at the surface of all the models significantly decreases
to a value less than $30$~km/sec.
Note that the observed radial velocity of Tycho G is $94 \pm 27$ km s$^{-1}$ by \cite{2004Natur.431.1069R} 
and $79 \pm 2$ km s$^{-1}$ by \cite{2009ApJ...701.1665K}, 
but the linear speed in our models are about $\sim 200$ km s$^{-1}$ (Table~\ref{tab_post_models}), 
corresponding to an inclination angle $\sin i \sim 0.4$.
Since the specific angular momentum $h_{\rm surface} \propto R_*^2 \Omega(R_*) \propto R_* v_{\rm rot}$, the surface rotation speed decreases while the post-impact remnant star is expanding.
For the rapidly evolving stars such as models A,B, and D, the rotation speed decreases to less than $10$~km/sec within 500 years.
After $1,000 - 1,500$~yrs, the stars start to contract, slowly increasing the surface rotation speed, 
but the surface rotation speeds of all models at 440 years are still below $30$~km s$^{-1}$. 
In addition, the surface rotation speed of models A, B, C, and D lie below $10$~km/sec at 440 years (Table~\ref{tab_440yrs}).
Although the surface rotation speed in our model E is still above the upper limit found by \cite{2009ApJ...701.1665K}, our progenitor models show that the post-impact remnant stars do not need to be fast rotators in the supersoft channel (WD+MS).
However, Tycho G could also be a stripped giant star that has lost most of its angular momentum during the SN Ia impact and then cooled \citep{2009ApJ...701.1665K}, or an expanded and slowed-down M dwarf as suggested by \cite{2012arXiv1209.1021W}.

\section{CONCLUSIONS}
We have investigated the post-impact evolution of remnant stars in the SDS for SNe~Ia via stellar evolution calculations. 
We examined six possible binary companion models in the mass-orbital-period space from \cite{2008ApJ...679.1390H} and one companion model in \cite{2012ApJ...750..151P}, 
and we performed three-dimensional hydrodynamics simulations using the setup in \cite{2012ApJ...750..151P}.
The post-impact evolution of the surviving stars was studied by using the stellar evolution code MESA, together with the reconstructed hydrostatic remnant star models.
It is found that the luminosity of post-impact remnant stars increases to 
$10-50 L_\odot$ after the supernova impact and increases to $\sim 100 L_\odot$ within a few thousand years, depending on the progenitor model.
Due to the energy deposition from the SN ejecta, the envelope of the post-impact remnant expands on its local thermal timescale ($\sim10^2-10^4$~yrs).
After this expansion, stars start to contract and release gravitational energy.  
The post-impact evolution is directly affected by the explosion energy since it is related to the amount of unbound mass after the 
SN impact and the amount and depth of energy deposited in the remnant star.
Among the calculated models, companion E in our simulation (see Table~\ref{tab_models}), 
which has a mass $M=1.44 M_\odot$, radius $R=4.57 R_\odot$, effective temperature $T_{\rm eff} = 5,737$~K, 
and luminosity $L=20.3 L_\odot$ after 440 years of the SN~Ia explosion,  
is closest to the observed properties of Tycho G as determined by \cite{2009ApJ...691....1G}
Although the fits are promising, the luminosity is twice as large as the value given by \cite{2009ApJ...691....1G}.    
Finally, by comparing the observed radial velocity to the linear speed in our progenitor models,
an inclination angle $\sin i \sim 0.4$ can be inferred.
The surface rotational speed thus implied ($\sim 10-20$ km~s$^{-1}$) approaches the low upper limit on the rotational speed of Tycho G found by \cite{2009ApJ...701.1665K}.
Our results provide some support for Tycho G as a possible progenitor candidate in the SDS 
and point to the need for further detailed studies of the SDS binary evolutionary channel.


\acknowledgments
KCP acknowledges Aaron Dotter, Bill Paxton, and Jing Luan for useful discussions about model loading in MESA. 
The simulations presented here were carried out using the NSF XSEDE Ranger system at the 
Texas Advanced Computing Center under allocation TG-AST040034N.  
FLASH was developed largely by the DOE-supported ASC/Alliances Center for Astrophysical Thermonuclear Flashes at the University of Chicago.  
This work was partially supported by NSF AST-0703950 to Northwestern University and by the Computational Science and Engineering (CSE) fellowship at the University of Illinois at Urbana-Champaign.
Analysis and visualization of FLASH data were completed using the analysis toolkit {\tt yt} 
\citep{2011ApJS..192....9T}.

\bibliography{ref}


\end{document}